\documentclass[12pt,preprint]{aastex}

\begin{document}

\title{{\it Spitzer Space Telescope} IRAC and MIPS Observations
of the Interacting Galaxies IC 2163 and NGC 2207: Clumpy Emission}

\author{Debra Meloy Elmegreen\altaffilmark{1},
Bruce G. Elmegreen\altaffilmark{2}, Michele
Kaufman\altaffilmark{3}, Kartik Sheth\altaffilmark{4}, Curtis
Struck\altaffilmark{5}, Magnus Thomasson\altaffilmark{6},and Elias
Brinks\altaffilmark{7}}

\altaffiltext{1}{Department of Physics \& Astronomy, Vassar
College, Poughkeepsie, NY 12604; elmegreen@vassar.edu}
\altaffiltext{2}{IBM Research Division, T.J. Watson Research
Center, P.O. Box 218, Yorktown Heights, NY 10598;
bge@watson.ibm.com} \altaffiltext{3}{Department of Physics and
Department of Astronomy, Ohio State University, 174 W. 18th
Avenue, Columbus, OH 43210; rallis@mps.ohio-state.edu}
\altaffiltext{4}{{Spitzer Space Center, Caltech, MS 220-6,
Pasadena, CA 91125; kartik@astro.caltech.edu} }
\altaffiltext{5}{{Department of Physics \& Astronomy, Iowa State
University, Ames, IA 50011; struck@iastate.edu} }
\altaffiltext{6}{{Onsala Space Observatory, S-439 92 Onsala,
Sweden; magnus@oso.chalmers.se} } \altaffiltext{7}{{University of
Hertfordshire, Centre for Astrophysics Research, College Lane,
Hatfield AL10~~9AB, United Kingdom; ebrinks@star.harts.ac.uk} }

\begin{abstract}
IC 2163 and NGC 2207 are interacting galaxies that have been well
studied at optical and radio wavelengths and simulated in
numerical models to reproduce the observed kinematics and
morphological features. Spitzer IRAC and MIPS observations
reported here show over 200 bright clumps from young star
complexes. The brightest IR clump is a morphologically peculiar
region of star formation in the western arm of NGC 2207. This
clump, which dominates the H$\alpha$ and radio continuum emission
from both galaxies, accounts for $\sim12$\% of the total 24$\mu$m
flux. Nearly half of the clumps are regularly spaced along some
filamentary structure, whether in the starburst oval of IC 2163 or
in the thin spiral arms of NGC 2207. This regularity appears to
influence the clump luminosity function, making it peaked at a
value nearly a factor of 10 above the completeness limit,
particularly in the starburst oval. This is unlike the optical
clusters inside the clumps, which have a luminosity function
consistent with the usual power law form. The giant IR clumps
presumably formed by gravitational instabilities in the compressed
gas of the oval and the spiral arms, whereas the individual
clusters formed by more chaotic processes, such as turbulence
compression, inside these larger-scale structures.
\end{abstract}
\keywords{galaxies: interacting --- galaxies: individual (NGC
2207) --- galaxies: individual (IC 2163) --- infrared: galaxies}

\section{Introduction}
IC 2163 and NGC 2207 are interacting galaxies at a distance of
35~Mpc (Elmegreen et al. 1995a, hereafter Paper I). IC 2163 has an
unusual bright oval of star formation shaped like an eyelid
(``ocular'' structure) at mid-radius that was predicted to be the
result of large-scale gaseous shocks in simulations of grazing
prograde encounters (Elmegreen et al. 1991; Sundin 1993). Several
ocular or post-ocular galaxies and their companions have been
observed and modeled in detail by our group (Kaufman et al. 1997;
Kaufman et al. 1999; Kaufman et al. 2002). The IC 2163/NGC 2207
pair was observed at optical (Elmegreen et al. 2000, hereafter
Paper III; 2001, hereafter Paper IV), millimeter (Thomasson 2004),
and radio wavelengths (Paper I; Elmegreen et al. 1995b, hereafter
Paper II). HST WFPC2 observations were used to measure the
properties of compact star clusters, many of which satisfy the
criteria for super star clusters (SSC; Paper IV), and to determine
the opacities of dust clouds in the foreground screen provided by
NGC 2207 (Paper IV). The HST observations also showed that the
clusters and stellar groupings in five major star-forming regions
are hierarchically distributed and that the fractal dimensions of
the star fields are comparable to the fractal dimension of
interstellar gas (Elmegreen \& Elmegreen 2001). Also, HST PC
images of the central region of NGC 2207 revealed what appears to
be acoustic turbulence (Elmegreen et al. 1998; Montenegro, Yuan \&
Elmegreen 1999; for other examples, see Martini et al. 2003).

Detailed numerical simulations have reproduced over two dozen
dynamical and structural details of the IC 2163/NGC 2207 pair
(Paper II; Paper III; Struck et al. 2005, hereafter Paper V). The
simulations suggest that NGC 2207 is moving prograde in the plane
of IC 2163, causing tidal forces that initially compressed and
stretched IC 2163 so that the stars were jarred into caustic
patterns and the gas was shocked into bright star-forming ridges.
Tidal forces in NGC 2207 were perpendicular to its disk, causing
perpendicular oscillations and possible warping (Paper II). There
is widespread high velocity dispersion in the HI gas in these
galaxies, commonly reaching $\sim50$ km s$^{-1}$ in otherwise
normal-looking regions. These dispersions have since been found in
all the ocular interactions (see references above) and in other
near collisions (e.g., Irwin 1994). Because of the extraordinarily
large HI clouds ($10^8$ M$_\odot$) that are also present (as
observed also in other close encounters -- Irwin 1994; Smith
1994), the observed velocity dispersions promoted the idea that
tidal dwarf galaxies can form because of super-size gravitational
instabilities in the gas (Elmegreen, Kaufman, \& Thomasson 1993;
Wetzstein \& Burkert 2005).  An even larger cloud located at the
tip of the far-western arm in NGC 2207 and containing $10^{8.9}$
M$_\odot$ appeared to be triggered by the interaction as well;
tidal forces can accumulate outer disk material at this point
(Mirabel, Dottori, \& Lutz 1992; Elmegreen et al. 1993; Duc,
Bournaud, \& Masset 2004).

NGC 2207 contains a peculiar region of intense emission at
H$\alpha$ and radio continuum wavelengths (hereafter called
Feature i, following the nomenclature in Paper III). Feature i is
located on a spiral arm in the western outer disk. Supernova
1999ec occurred just south of this region, near two other small IR
clumps (see Fig. \ref{fig:examples} below). The H$\alpha$ luminosity of
Feature i, spanning 6 arcsec in diameter, is equivalent to that of
30 Dor in the Large Magellanic Cloud. The radio continuum emission
comes from a $\sim1$ arcsec core with $\sim300$ times the
luminosity of Cas A at 6-20 cm (Vila et al. 1990), and an extended
component ($\sim2$ kpc diameter) with $\sim1500$ times the
luminosity of Cas A (Paper II). HST reveals an unusual dust cloud
elongated for $300$ pc perpendicular to the local spiral arm and
containing $>10^6$ M$_\odot$ of gas with $>3$ magnitudes of
optical extinction (Paper IV). HST also shows a red V-shaped
feature extending for $\sim500$ pc in the opposite direction to
the dust cloud (Paper III). Between the dust cloud and the red V
is a very red region that may contain several massive clusters.
Perhaps Feature i is a young star cluster that formed with such a
high density that the central stars combined into a black hole, as
in models by Ebisuzaki et al. (2001).

Spitzer Space Telescope (SST) observations of other galaxies
reveal old and young stars at the shortest wavelength ($3.6\mu$m)
in the Infrared Array Camera (IRAC, see Fazio et al. 2004), and
PAH emission and heated dust continuum at the longer IRAC
wavelengths ($4.5\mu$m, $5.8\mu$m, and $8.0\mu$m; e.g., Helou et
al. 2004; Willner et al. 2004). The 8$\mu$m band in IRAC and the
longer wavelength bands in the Multiband Imaging Photometer (MIPS;
Rieke et al. 2004) tend to show more filamentary structures, such
as those seen in M81 (Gordon et al. 2004a).

Here we present IRAC and MIPS observations of IC 2163/NGC 2207
with the SST in order to consider the distribution and properties
of clumpy emission at IR wavelengths. The HST data showed the
presence of many super star clusters (SSCs), with the most massive
having comparable luminosities to the most massive clusters in the
Antennae galaxies (Zhang \& Fall 1999). This is peculiar for IC
2163/NGC 2207 because the total number of clusters is 100 times
smaller than in the Antennae. Generally, larger samples of
clusters have brighter brightest members, with a near-linear
relation between brightest luminosity and number of members for an
$n(L)dL\propto L^{-2}dL$ luminosity ($L$) distribution function
(Whitmore 2002). The similarity of the brightest members in IC
2163/NGC 2207 compared with the much more prolific Antennae galaxy
suggests that the cluster luminosity function in IC 2163/NGC 2207
is either a flatter power law than what is seen in the Antennae,
or it is peaked at the high mass end. The most massive clusters
are in the eyelid part of IC 2163, which, according to our models,
is a galactic shock front caused by the sudden radial motion of
disk gas responding to tidal torques. Many other clusters have
formed in filamentary shocks as well.

The SST images are not resolved well enough to separate individual
star clusters within a clump. The clumps that can be observed are
the equivalent of star complexes, which are collections of OB
associations and dense clusters ranging in size from several
hundred parsecs to a kpc (Efremov 1995). In contrast, our HST
observations gave luminosities for individual clusters. We compare
these two data sets here. The extended IR emission that was
observed with SST will be the subject of a forthcoming paper
(Kaufman et al. 2005).

\section{Data and Overall Emission}
Observations were made with the Spitzer Infrared Array Camera on
February 22, 2005 and the Multiband Imaging Photometer on March
11, 2005.  For the IRAC observations at 3.6, 4.5, 5.8, and
8$\mu$m, the full array readout mode was used, with high dynamic
range. The frame time was 30 sec, using a cycling dither pattern
with 9 positions.  The images have a scale of 1.2 arcsec
pixel$^{-1}$, corresponding to $\sim200$ pc at 35~Mpc. For the
MIPS observations, the large field size was used for $24\mu$m and
160$\mu$m, and the small field size for 70 $\mu$m. For 24$\mu$m,
the exposure time was 30 sec. For 70$\mu$m, a raster map was made
with a 3x3 raster and a step size of 1/2, with a 10 sec exposure
time. For 160$\mu$m, 4 cycles were used, with 10 sec per cycle.
The image scales are 2.45 arcsec pixel$^{-1}$, 4.0 arcsec
pixel$^{-1}$, and 8.0 arcsec pixel$^{-1}$ for the $24\mu$m,
$70\mu$m, and 160$\mu$m bands, respectively. The corresponding
point-spread functions are 2.0 - 2.4 arcsec (FWHM) for the IRAC
images and 5.9 arcsec, 18 arcsec, and 40 arcsec, respectively, for
the MIPS bands.

\begin{figure}
\epsscale{1.0} \plotone{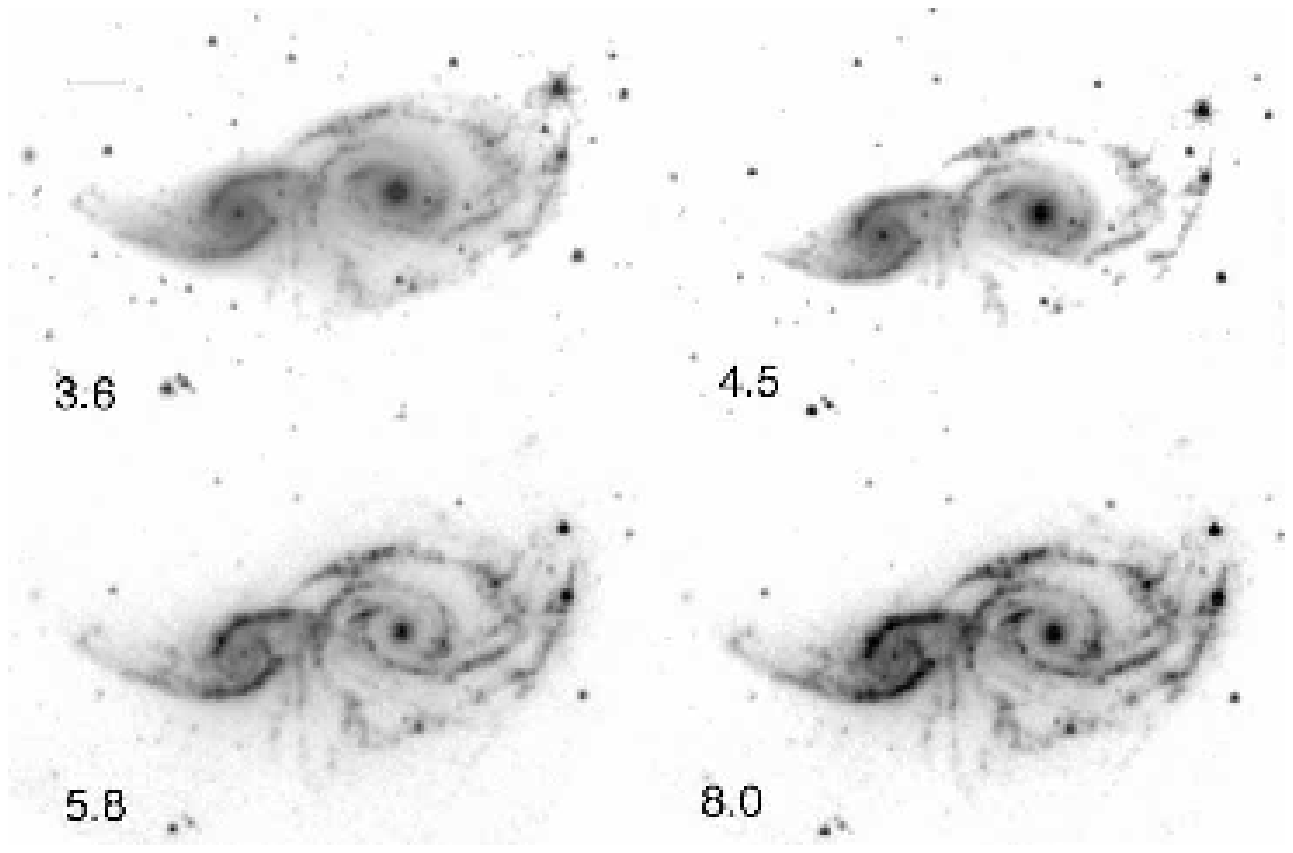} \plotone{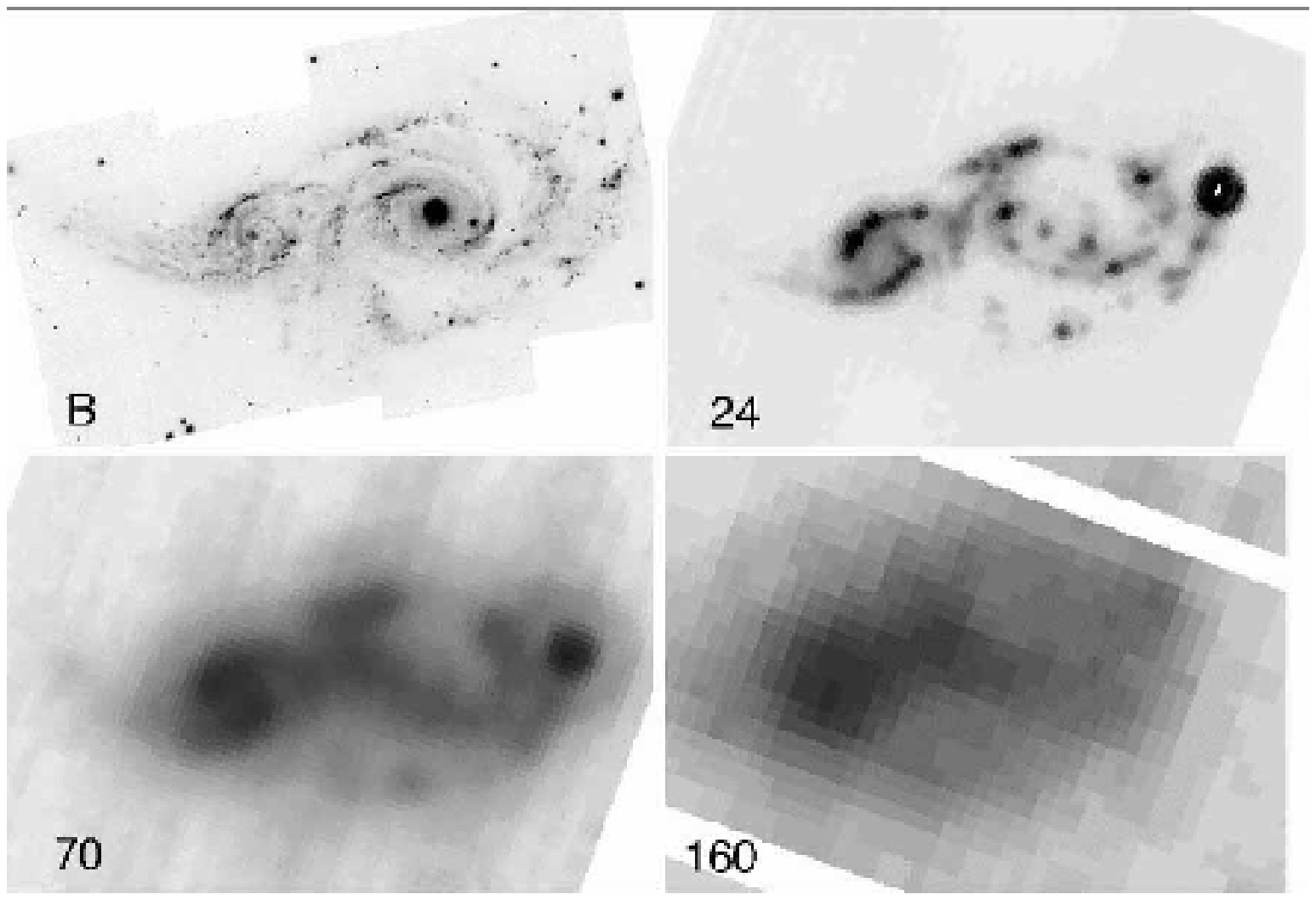}\caption{IRAC
images (top) and MIPS images with wavelengths indicated, all
plotted on a logarithmic intensity gray scale. The HST WFPC2
B-band image is also shown. The line in the upper left shows a
scale of 30 arcsec (5.1 kpc). (Image degraded for astroph)} \label{fig:mosaic}\end{figure}

The Basic Calibrated Data images are shown in Figure
\ref{fig:mosaic}, with each band represented separately for IRAC
and MIPS.  For comparison, a B-band HST WFPC2 image is also shown
(from Paper IV).  The line in the upper left shows a scale of 30
arcsec, which corresponds to 5100 pc at the distance of 35~Mpc.
The clumps and spiral arms in the IRAC bands match the HST optical
features well. There is a semi-circular shell 1600 pc in diameter
in the southern arm of NGC 2207 (marked in Figure \ref{fig:color}
below) that is not obvious in the optical images but partially
coincides with an H$\alpha$ feature (Fig. \ref{fig:hairac4} below)
and a dust arc.

Feature i stands out in all IRAC and MIPS images as an intense
source in the western outermost arm of NGC 2207. The central 2
pixels of Feature i are saturated in the 24$\mu$m image. The
eyelid region of IC 2163 is also prominent in all IRAC and MIPS
bands; it dominates the 160$\mu$m image, although it is
unresolved. The MIPS 70$\mu$m image is similar to the 20 cm radio
continuum image (shown in Paper I).  A detailed comparison between
MIPS, HI, and radio continuum emissions will be made in a second
paper (Kaufman et al. 2005).

Figure \ref{fig:color} shows a color composite image of the four
IRAC bands, with colors ranging from blue for the short wavelength
band to red for the long wavelength band. The prominent features
discussed in this paper are identified. The spiral arms in NGC
2207 are clearly segmented into regularly spaced beads of
emission. The same is true for the oval in IC 2163, although the
individual beads are difficult to see with the contrast level in
this color image (they show up better in Figs. \ref{fig:divide}
and \ref{fig:beads} below).

Figure \ref{fig:hairac4} is an overlay of 8$\mu$m contours on an
H$\alpha$ emission image (from the ground-based
continuum-subtracted H$\alpha$ image in Paper IV).  The bright
8$\mu$m clumps match the HII regions well. A generally good match
between $8\mu$m and H$\alpha$ was also found for NGC 300 by Helou
et al. (2004). Kennicutt (2005) noted that 95\% of the infrared
clumps in nearby galaxies of the Spitzer Infrared Nearby Galaxy
Survey (SINGS -- Kennicutt et al. 2003) have associated HII
regions, and that this implies star-forming regions take less than
1 Myr to become optically visible.

A map made by dividing the 3.6$\mu$m image by the 4.5$\mu$m image
is shown in Figure \ref{fig:divide}. The nuclei disappear and the
star-forming clumps become more prominent. Evidently, the
[3.6]/[4.5] ratio is about constant everywhere outside the clumps.
This implies that the mixture of field stars and diffuse hot dust
emission is uniform outside the clumps. Inside the clumps, the
[3.6]/[4.5] ratio is low (black in the image). This low ratio may be the
result of an excess of H Br$\alpha$ at $4.5\mu$m from ionization
inside the clumps (Churchwell et al. 2004; Whitney et al. 2004);
the hot and bright stars which contribute to this
ionization are not prominent at either $3.6\mu$m or $4.5\mu$m.  Figure
\ref{fig:cprofile} shows scans through both nuclei out to the
galaxy edges in 3.6$\mu$m and 4.5$\mu$m and in the ratio
[3.6]/[4.5]. The uniformity of the interclump [3.6]/[4.5] ratio is
also evident here.  Clumps at the far right of the scan (west in
the image) break this uniformity with low values of [3.6]/[4.5].

The total flux densities in IRAC and MIPS were determined for the
combined galaxies, the eyelid and interior region of IC 2163, and
Feature i. We used the {\it imstat} routine in the Image Reduction
and Analysis Facility (IRAF) to measure the total flux inside a
box with a fixed size for each region. Backgrounds determined from
neighboring boxes were subtracted.  The results are given in Table
1. For Feature i, the box size was $24^{\prime\prime}$, which is 3
times the MIPS $160\mu$m pixel scale.

\section{Clumpy Emission}
\subsection{IRAC colors}
\label{sect:iraccolors} Photometry for 225 IRAC clumps, identified
by eye in the 3.6$\mu$m image, was done in all possible passbands
using {\it phot} in IRAF. All the measured clumps are circled in
Figure \ref{fig:id}; stars are denoted by squares. The clumps also
show up in the other passbands (Fig \ref{fig:mosaic}) but the
longer wavelengths show the more filamentary, gaseous
structures. The brightest clumps in the figure are numbered and
Feature i is indicated; their properties are listed in Table 2,
including H$\alpha$ flux. The positions and IRAC fluxes of all the
measured clumps are presented in Table 3.

The photometry was performed using a fixed aperture 3 pixels in
radius with background determined from an annulus 5 pixels in
width, separated from the aperture by 5 pixels.  The IRAC counts
in MJy/ster were converted to flux densities by multiplying by the
areas of the pixels, and then converted to magnitudes using the
zeropoints given in the online IRAC tables of the Data
Handbook\footnote{http://ssc.spitzer.caltech.edu/irac/dh, Chapter
5}. The photometric accuracy of the IRAC images is better than 0.1
mag. We find that the measurement uncertainties are 0.2 to 0.5
mags for different regions because of varying sky backgrounds and
variations within the clumps. The faintest clump was 18.7 mag at
3.6$\mu$m, which corresponds to an absolute magnitude of $-14.0$
at a distance of 35~Mpc.

Color-magnitude diagrams for all measured IRAC clumps are shown on
the left in Figure \ref{fig:cm}. Different symbols correspond to
different locations in the galaxies: inner regions (triangles),
eyelid or middle (squares), and outer regions (circles). Open
symbols are for IC 2163 and filled symbols are for NGC 2207. The
rms scatter in color increases as the clumps get fainter. We
judged that the color errors start to be significant for [3.6] and
[4.5] magnitudes fainter than 16, [5.8] magnitudes fainter than
14, and [8.0] magnitudes fainter than 12.5. Complete removal of
any object with at least one passband fainter than these limits
led to 116 remaining clumps, which we will refer to as the
low-noise group; the other clumps will be referred to as the
high-noise group. The color-magnitude diagrams for the 116
low-noise clumps are shown on the right of Figure \ref{fig:cm}.
The brightest object in all diagrams is Feature i in NGC 2207. The
next brightest objects are in the eyelid of IC 2163, most of which
are 1-2 mag brighter at 3.6$\mu$m than any other clump.

Figures \ref{fig:cc123} and \ref{fig:cc234} show color-color plots
of the 116 low-noise clumps. They all cluster in a narrow region
on these plots, indicating some uniformity to the properties of
the interstellar medium.  There are no obvious trends between
color and location. In Figure \ref{fig:cc123}, the colors of dust
(Li \& Draine 2001) and G0-M5 stars (Jones et al. 2005) are shown
by a plus symbol and diamond, respectively (from Smith et al.
2005; Whitney et al. 2004). In Figure \ref{fig:cc234}, the plus
symbol shows the dust again, but the G0-M5 stars would be off
scale at [4.5]-[5.8]$\sim0.1$ and [5.8]-[8.0]$\sim0.1$. The nuclei
of both galaxies are shown by X symbols in both figures; they have
essentially the same colors as G0-M5 stars (Jones et al. 2005).
The [3.6]-[4.5] colors of the clumps are redder than G0-M5 stars
by $\sim0.1-0.3$ mag, signifying populations with average ages
younger than A0V stars (e.g., Pahre et al. 2004a). The [4.5]-[5.8]
colors are redder than G0-M5 stars by $\sim1.5-3$ mag, again
indicating young ages.

Several studies have considered whether IR emission from a given
region is starlight-dominated, PAH-dominated, or blackbody
emission in certain passbands. Engelbracht et al. (2005) examined
IR spectra, IRAC, and MIPS data of spiral galaxies, compared them
with stellar evolution models, and showed a clear division between
sources containing PAH emission and sources not containing PAH
emission based on the ratio of starlight subtracted 4.5$\mu$m flux
density to 8$\mu$m flux density; a ratio less than 0.05
corresponds to the presence of PAH emission. They assumed that the
3.6$\mu$m emission is dominated by old starlight (e.g., Helou
2004). Following their method, we subtracted a scaled 3.6$\mu$m
emission (with a factor 0.57 which they based on Starburst 99
models from Leitherer et al. 1999) from the 4.5$\mu$m emission for
each clump; the ratio of this reduced 4.5$\mu$m emission to the
8$\mu$m emission fell below 0.05 for 215 out of the 225 clumps,
indicating that these clumps should contain PAH emission. Such PAH
emission is expected to contribute mostly to 5.8 $\mu$m and
8$\mu$m bands (e.g. Helou et al. 2004, Wang et al. 2004). Sajina,
Lacy, \& Scott (2005) showed that PAH's are located in a different
region of $\log [S_8/S_{4.5}]$ versus $\log [S_{5.8}/S_{3.6}]$
than starlight (see their Fig. 6). Our clumps are almost all in
the range of $\log\left(S_8/S_{4.5}\right)=0$ to 1,
$\log\left(S_{5.8}/S_{3.6}\right)=0.7$ to 1.5 for these flux
ratios, which falls within the PAH regime of the diagram. Also,
our clump color distribution for [5.8]-[8] peaks between 1.6 and
2.2, which is consistent with PAH-dominated emission at these
wavelengths, as modeled by Li \& Draine (2001).

In other SST observations of interacting galaxies, Arp 107, at a
distance of 138~Mpc, was found to have numerous clumps in a
ring-like distribution with a 1.3 mag azimuthal variation of
[4.5]-[5.8] colors indicating a stellar age gradient, as matched
by numerical simulations (Smith et al. 2005). The clumps studied
in Arp 107 were divided into brighter and fainter ones. The
brighter clumps had slightly redder [4.5]-[5.8] and [5.8]-[8.0]
colors (by about 0.5 mag) than the fainter clumps. Smith et al.
interpret the redder clumps as having a higher proportion of young
to old stars. Our clumps on average have about the same
[3.6]-[4.5] colors as all of their clumps, $\sim$-0.2 to 0.3, with
colors that are about 0.5 mag redder than their bright clumps in
[4.5]-[5.8] and similar to their bright clumps in [5.8]-[8.0].
These redder colors suggest that our regions are slightly younger
on average than theirs.

Spitzer IRAC observations of the Antennae galaxy pair NGC
4038/4039, at a distance of 21~Mpc, reveal clumpy star formation
and extensive PAH emission (Wang et al. 2004).  Their average
[5.8]-[8.0] clump color is $\sim 1.8$ mag, which is consistent
with our clump colors. M51 also has prominent star-forming clumps
and bright filaments in SST images (Calzetti et al. 2005). IRAC
and MIPS observations of M81 (Gordon et al. 2004a) and M31 (Gordon
et al. 2004b) show that the prominent clumpy UV, 8$\mu$m, and
24$\mu$m features correspond well to one another.

The alignment between IRAS clumps and H$\alpha$ emission in
IC2163/NGC 2207, the color-color distribution of the IRAC clumps,
and the likely presence of PAH emission from them, all suggest
that these regions are relatively young. Precise age estimates are
not possible without better calibrations of the relative
contributions to the IRAC bands from young and old stars, dust,
free-free emission, PAHs, and other possible emission sources.
Nevertheless, the young ages found here are consistent with the
ages derived by more conventional techniques using HST data, which
are up to several $\times10^7$ years (Paper IV).

\subsection{Sample Comparisons with HST emission}

Each IRAC clump is composed of numerous smaller clusters and
associations that are individually visible with HST. The super
star clusters found with HST, for example, are inside the IRAC
clumps, although these IRAC clumps do not appear to be more
luminous or special in any way.

Figure \ref{fig:examples} shows two IRAC star forming regions in
detail; Feature i is on the left and clump IR 9 from the eyelid
region is on the right. The contours are $8\mu$m emission at 2.2
arcsec resolution and the gray scale is the optical emission at
0.13 arcsec resolution. In the vicinity of Feature i, there are
two $\sim10^{4.5}$ M$_\odot$ super star clusters (SSC) from the
catalog of Paper IV, many other clusters and associations, four
IRAC sources from Table 3, an optical supernova source, and a dust
streak to the southwest of a very red cluster near the center of
the large $8\mu$m source. In the vicinity of IR 9, there are 2
super star clusters and 2 other IRAC sources.

\subsection{Comparison with H$\alpha$ emission}
The luminosities of HII regions were calculated in Paper IV based
on ground-based H$\alpha$ images.  The H$\alpha$ resolution is
similar to the IRAC resolution. For individual clumps there is a
correlation between the H$\alpha$ luminosity, which ranges from
$5\times10^4$ L$_\odot$ to $4\times10^6$ L$_\odot$, and the
passband-combined IRAC luminosity, which ranges from
$2.5\times10^6$ L$_\odot$ to $2.5\times10^8$ L$_\odot$ (using
$L_\odot=4\times10^{33}$ erg s$^{-1}$). There is a scatter around
this correlation of $\sim0.5$ mag, which is most likely the result
of slight age and extinction differences. The luminosities are
compared in Figure \ref{fig:halpha}. The brightest feature in both
passbands is again Feature i.  For the individual clumps, the
average ratio of H$\alpha$ to 3.6$\mu$m is $0.052\pm0.062$. For
Arp 107, this ratio averages 0.04 (Smith et al. 2005).  For
H$\alpha/8\mu$m, our average ratio is $0.005\pm0.006$ and for Arp
107, the ratio ranges from 0.006 to 0.05.

For the combined galaxies, the ratio of H$\alpha$ to 3.6$\mu$m
luminosity is 0.009.  This ratio is slightly less than that in Arp
107, where it is 0.012 for the large galaxy and 0.025 for the
small galaxy (Smith et al. 2005). It is also slightly less than in
NGC 7331, where it is 0.013 (Regan et al. 2004). The ratio of
H$\alpha$ to $8\mu$m for the combined IC 2163/NGC 2207 pair is
0.007, which is slightly less than 0.0083 and 0.025 for Arp 107,
and 0.030 for NGC 7331.

\subsection{Spectral energy distributions}
The IRAC SEDs for the clumps in Table 2 are shown in Figure
\ref{fig:sedhubble}, with the log of the wavelength times flux
density, $\lambda F_{\lambda}$, versus the log of the wavelength.
The B, V, and I flux densities from HST are also plotted. They
were measured in the HST images using apertures 36 pixels in
radius to correspond to the apertures used for the IRAC images.
Some of the SEDs dip down toward the blue, presumably because of
extinction in these optical bands (see Paper IV). All of the
clumps show a dip at 4.5$\mu$m.

The brightest clump is Feature i, which is represented twice in
Figure \ref{fig:sedhubble}: once as an IRAC flux like the other
clumps and again as a flux from all passbands inside a
$24^{\prime\prime}$ box, including MIPS images. The line with open
circles indicates the IRAC values using the usual
$1.2^{\prime\prime}$ pixel size of this instrument, with photometry
done in an aperture of 3 pixel radius, as for the other clumps.
The flux inside the $24^{\prime\prime}$ box is indicated by a
dotted line and x-marks; it has the background subtracted as
determined from neighboring boxes. The two curves for Feature i
are similar at short wavelengths because the 3-pixel central
source dominates the emission inside the $24^{\prime\prime}$ box.
The combined light from IC 2163 and NGC 2207 is shown in Figure
\ref{fig:sedhubble} as the long solid line. The eyelid plus
interior of IC 2163 is shown as a long dashed line.

The integrated SEDs from other galaxies observed in the SINGS
survey are similar to the integrated SED in Figure
\ref{fig:sedhubble}. Dale et al. (2005) show SEDs for galaxies
spanning a wide range of Hubble types. Our combined-galaxy SED
resembles the integrated SEDs for the Seyfert I galaxy NGC 1566
(SABbc), the HII galaxy NGC 2403 (SABcd), and the LINER galaxies
NGC 1097 (SBb), NGC 7552, and NGC 3190 (SAap). These galaxies all
have a dip at 4.5$\mu$m, a local peak at 5.8 or 8$\mu$m, a trough
at $24\mu$m, and a rise to $70\mu$m or $160\mu$m, like IC2163/NGC
2207 combined. There are also other galaxies in Dale et al. (2005)
that are qualitatively similar; many of them are LINERs or
Seyferts.

Theoretical curves in Dale et al. (2005) fit the region from
$24\mu$m to $160\mu$m with continuum emission from dust. We
determined dust temperatures from MIPS data for the combined
galaxies, the eyelid and interior regions of IC 2163, and Feature
i. They are based on the flux densities in Table 1. Because there
can be different dust components contributing to the short and
long wavelength bands of MIPS, we calculated blackbody
temperatures from both the $24\mu$m/$70\mu$m ratio and the
$70\mu$m/$160\mu$m ratio. Dust temperatures were not determined
for individual clumps because of the lack of resolution in the
long wavelength bands.  For these three regions, the dust
temperatures from the short wavelength ratios were approximately
50K, 52K, and 58K, respectively. Similarly, the dust temperatures
from the long wavelength flux ratios were 25K, 24K, and 36K,
respectively. The uncertainties in our flux densities are about
20\% based on the MIPS Data Handbook\footnote{http://ssc.spitzer.caltech.edu/mips/dh/mipsdatahandbook3.1.pdf}.
This would give a temperature range of $\pm$3K for the above
measurements. Thus, the temperatures for the combined galaxies and
the eyelid are essentially the same, but the temperature in
Feature i is slightly warmer. This is also evident from the SED in
Figure \ref{fig:sedhubble}, which shows a downturn from $70\mu$m
to $160\mu$m for Feature i, unlike the other SEDs.

\subsection{IRAC Luminosity Functions}

Clump luminosities were determined from the IRAC bands. The MIPS
bands contain diffuse ISM emission that is unrelated to star
complexes (Haas et al. 2002), and they are also poorer resolution,
so we do not include them here. For the IRAC bands we integrated
the flux density over wavelength,
\begin{equation}
L_{\rm IRAC}\left({\rm erg\;\;s}^{-1}\right)= 4\pi D^2
10^{-23}\sum_{i=1}^4 F_i\Delta\lambda_i,
\end{equation}
where the bandwidths $\Delta\lambda$ for the 4 bands are
$0.75\mu$m, $1.015\mu$m, $1.425\mu$m, and
$2.905\mu$m\footnote{http://ssc.spitzer.caltech.edu/documents/som/irac60.pdf,
page 82}; $i=1$ corresponds to $3.6\mu$m, $i=2$ corresponds to
$4.5\mu$m, etc., and $F_i$ is the flux density in Jy per micron
($=10^{-4}F_{\nu}c/\lambda^2$ with cgs units for speed of light
$c$ and wavelength $\lambda$). For this equation, $D$ is the
distance of 35~Mpc, evaluated in cm.

Figure \ref{fig:gaush} shows the IRAC luminosity functions for the
three main regions of each galaxy (inner, middle, and outer, as in
Figs. \ref{fig:cm}-\ref{fig:cc234}). The bar-chart histograms are
counts in bins of width 0.5 in the log of the luminosity (in
$L_\odot$) for the low-noise clumps. The open squares are counts
for all the clumps, including the high-noise clumps.  The x-marks
on the abscissae are the sensitivity limits, $10^{6.57}$ L$_\odot$,
defined by an integration over the separate limits mentioned in
reference to Figure \ref{fig:cm}: 16th mag for [3.6] and [4.5],
14th mag for [5.8], and 12.5 mag for [8].

The smooth curves in Figure \ref{fig:gaush} calculate the
luminosity distribution in a different way, this time including
both the low-noise group (116 clumps, plotted with solid-line
curves) and the combined low and high-noise groups ($116+94$
clumps, plotted with dashed curves). In this method, we add a
Gaussian distribution for each clump with an area equal to unity
and a dispersion equal to the uncertainty in the luminosity
obtained from the uncertainty in the photometry measurement (as
calculated by {\it phot} in IRAF). This error is low for the
brighter clumps (0.16 of a magnitude at log L$_\odot=7.9$) and
relatively high for the faintest clumps (0.3 of a magnitude at log
L$_\odot=6.57$, which is the limiting sensitivity). In this way,
the fainter clumps can be included with their high uncertainties.
The resultant summed distributions were then multiplied by 4 to
give them a height in Figure \ref{fig:gaush} comparable to the
histogram.  The Gaussian-sum distributions resemble the binned
distributions for the low-noise clumps (the dotted histograms),
and extend to lower luminosity for the high-noise clumps,
resembling the counts shown by open squares. The dots in the
Gaussian-sum curves represent the centers of the bins used for
counting.

The combined luminosity functions for all of the IRAC sources are
presented in Figure \ref{fig:gausa}. The panel on the left is for
the sum of the IRAC bands, and the panel on the right is for the
3.6 $\mu$m band only, calculated as $\Delta\lambda_{3.6}F_{3.6}$
and converted to cgs. As in the previous figure, the solid curves
are for the low-noise sources and the dashed curves are for all the
sources including those with high-noise. Dots on the curves
denote the centers of the counting bins. Feature i is the single
peak on the right in the left panel; it appears only in the
combined luminosity function. The open squares are counts for all
the sources in intervals of 0.5 mag. The two sensitivity limits
are shown as x-marks on the abscissae.

The combined IRAC luminosity function (Fig. \ref{fig:gausa}) is
steeper than the usual power law with a slope of $-1$ on such a
log-log plot. In Paper IV we also found that the HII region
luminosity function has a steep drop-off at the upper end.

A high fraction of clumps are in thin filaments, including the
whole eyelid region and many of the arcs of spiral arms. In these
regions, the complexes look like regular beads on a string (e.g.,
Figs. \ref{fig:color} and \ref{fig:divide}). Figure
\ref{fig:beads} illustrates several of the regions where the
clumps appear evenly spaced. The 8$\mu$m logarithmic image is
shown because this band has a particularly good correspondence
with star-forming regions, as noted earlier. The top left of
Figure \ref{fig:beads} shows the whole image, with boxes
indicating the strings of clumps that are shown enlarged below the
galaxies. There are other smaller strings evident in the figure
too. This beading phenomenon is fairly common here: 110 out of the
225 measured IRAC clumps are in more-or-less regular strings of
three or more members. The scale of 0.5 arcmin indicated in the
figure is for the expanded images; it corresponds to $\sim 5$ kpc.
The average spacing of the clumps in strings is $\sim 2-3$ kpc,
and the typical clump diameter is $\sim 0.5-1$ kpc. Figure
\ref{fig:beadhis} shows histograms of the clump luminosities in
each of the strings in Figure \ref{fig:beads}, including Feature i
in string f.

\subsection{Model Luminosity Functions} \label{sect:models}

Random power-law luminosity functions were made for comparison
with the galaxy data. The intrinsic function ranged between the
observed lower limit, $10^{6.57}$ L$_\odot$, and an arbitrary
upper limit of $10^9$ L$_\odot$, and was of the form
$n(L)dL\propto L^{-2}dL$, as typically observed (corresponding to
a slope of -1 when binning in $\log L$, as in Figs.
\ref{fig:gaush} and \ref{fig:gausa}). To make a Gaussian
contribution to the distribution function for each model clump,
i.e., having a width given by the uncertainty in the luminosity
and an area equal to unity, the model noise for each model
luminosity was determined from the empirical relationship between
the observed luminosity and the observed uncertainty in luminosity
(which ultimately came from the magnitudes and rms errors in
magnitude given by {\it phot}). Figure \ref{fig:gausm} shows the
resultant model distribution functions. The left column of panels
in Figure \ref{fig:gausm} is for three random trials with clump
numbers of $10^4$, $10^3$, and $10^2$. The top two panels on the
right are for different random trials with 116 clumps each, as in
the low-noise distribution functions for our combined galaxies.
The bottom right panel shows 10 random trials (shifted
successively upward for clarity) with only $N=11$ clumps, which is
the number observed in the eyelid region. Each curve is normalized
to have the same area; dots show the sampling interval for the
histograms.

Figure \ref{fig:gausm} suggests that the single bright clump found
at high luminosity in the galaxy data, which is Feature i, is not
unusual when the number of clumps brighter than $10^{6.57}$
L$_\odot$ in IRAC is as small as 116. The distribution functions
at high luminosity are ragged in our models, commonly giving a
single bright clump somewhat removed from the next brightest
clump.  We also see from Figure \ref{fig:gausm} that all of the
model functions with total numbers exceeding $\sim100$ show traces
of the intrinsic power law distribution extending down to the
sensitivity limit. The gradual drop-off below this limit is from
the finite Gaussian widths. Even random models with very low
numbers of complexes, 11 in these cases, have peaks in the
distribution functions close to the sensitivity limit. This
differs from the observed distributions for clumps in some
strings, particularly in the eyelid region. Secondary peaks may
occur at larger luminosities because of low-number statistics, as
in Figure \ref{fig:gausm}, but in the models, there is always a
prominent peak close to the limit.

The random-clump models reinforce our impression from Figures
\ref{fig:gaush} and \ref{fig:beadhis}: the distribution function
for the eyelid region of IC 2163 drops off too fast at low
luminosity to be the result of an intrinsic power law that extends
below the sensitivity limit.  The same may be true for other clump
strings, but the statistical sampling is too poor to be sure.
Also, the combined luminosity function for all 116 bright clumps
(Fig. \ref{fig:gausa}) appears to drop off too fast at
intermediate-to-high luminosity to be derived from an intrinsic
$L^{-2}dL$ function unless feature i is considered a member of the
sample.  Indeed, the IRAC luminosity of Feature i is not so
unusual for an $L^{-2}$ function: most of the models with 100 or
more clumps had the equivalent of a stand-out Feature i. As noted
before, there are other unusual aspects of Feature i as evident
from its peculiar morphology, but apparently not the mass or
luminosity alone.

Feature i would appear more unusual if the intrinsic luminosity
function fell significantly more rapidly than the assumed
$L^{-2}dL$ power law at high luminosity. If the distribution
function is $P(L)dL$, then the probability of finding an object
with $L>X$ is $\int_X^\infty P(L)dL/\int_{L_{\rm min}}^\infty
P(L)dL$. For $P(L)\propto L^{-2}$ as above, the probability of
finding a source with $L>10^{8.25}$ L$_\odot$ when $L_{\rm
min}=10^{6.5}$ L$_\odot$ is 1.8\%, so among 116 objects, two
Feature-i like sources might be expected, as discussed above. If
$P(L)$ is log-normal like the halo globular clusters,
$P(\xi)\propto \exp\left(\xi-\xi_0\right)^2/2\sigma^2$ for
$\xi=\log_{10}(L)$ with $\xi_0=7.4$ at the peak, $\sigma=0.5$, and
$\xi_0>>\sigma$ (cf. Fig. \ref{fig:gausa}), then the probability
that a source is more than $X$ times $\sigma$ brighter than the
peak is $1-{\rm erf}\left(X/2^{1/2}\right)$. Taking $X=2$ as in
Figure \ref{fig:gausa} gives a probability of 4.6\%, which amounts
to five Feature-i like sources among 116 total objects.  In either
case, Feature i does not appear to be unusual considering the
other bright clusters in these galaxies.

One explanation for the lack of a power law luminosity function in
the eyelid region is that the complexes could be so old that
fading and evaporation have depleted the low-mass members. This is
unlikely because the eyelid is a transiently compressed region
triggered by the recent interaction.  The clumps also appear to be
young because of their H$\alpha$ emission and IRAC colors (Sect.
\ref{sect:iraccolors}). In that case, we expect little disruption
and fading during their short lives, and a resulting luminosity
function for the brightest clumps that resembles the initial clump
mass function.

Another explanation for the peaked luminosity function is that the
clumps in the strings formed by regular gravitational
instabilities which gave them a characteristic size and mass. This
differs from the usual model where small clusters form in a
turbulent gas with hierarchical structure and an $M^{-2}dM$ power
law (Elmegreen \& Efremov 1997). The clump separations observed
here are about the same as those observed for star complexes in
the spiral arms of many grand design galaxies (Elmegreen \&
Elmegreen 1983). The formation process of these complexes is most
likely a gravitational instability in the spiral shock, which has
locally low shear and a high enough density to be unstable during
the arm residence time (Elmegreen 1994; see simulations in Kim \&
Ostriker 2002).  In this explanation, what makes the IRAC
luminosity function unusual (compared to the expected power-law)
is that IRAC measures whole star complexes and not individual
(small) clusters. The luminosity functions of young beads on a
string in grand design spiral galaxies could have the same peaked
shape. This is in contrast to the overall distribution of HII
complexes in galaxies, which have luminosity functions consistent
with a power law (Elmegreen \& Salzer 1999).

The most peculiar IRAC luminosity function in Figure
\ref{fig:gaush} is for the clumps in the eyelid region. We
measured the V and I band magnitudes of 165 individual star
clusters in this region (to deeper limits than in Paper IV) using
our HST data and 3-pixel apertures for the photometry. Clusters
could be distinguished from stars by their positions on a V-I
versus I-band color-magnitude diagram and by the relatively high
cluster luminosities.  The resulting cluster I-band luminosity
function is shown as the histogram in Figure \ref{fig:hst}. The
luminosity function of the sum of the cluster luminosities in each
SSC clump along the eyelid is shown by the dashed line. These are
the luminosities for emission only in the I-band filter and are
therefore less than the bolometric cluster luminosities. The
dotted line has a slope of $-1$, which is the typical power law
slope for clusters on a log-log plot. The high-luminosity part of
the cluster function does not follow any clear power law, but it
is close enough to the typical slope to be considered normal. The
peak in the histogram is close to the faintness limit where
measurement errors become important and fainter clusters start to
be lost in the noise.  The summed-luminosity function is
significantly above the detection limit and is peaked at a value
of $10^{3.5}$ L$_\odot$; this shape supports the IRAC observation
that the luminosity function for large clumps in the eyelid is not
a power law.

The eyelid luminosity function for IRAC clumps in Figures
\ref{fig:gaush} and \ref{fig:beadhis}, combined with the eyelid
luminosity function for individual HST clusters in Figure
\ref{fig:hst}, suggest that only the star complexes, representing
the blended large-scale regions of many young clusters, might be
considered to have a characteristic luminosity and a peaked
luminosity function. Individual clusters which comprise these
complexes appear to be more scale-free.

The peaked luminosity functions for IRAC clumps in the eyelid
region of IC 2163 and in several of the beaded strings of NGC 2207
are analogous to the inferred peaked initial mass function for
bright clusters in the B region of starburst galaxy M82 (de Grijs,
Parmentier, \& Lamers 2005). The M82 region is much older ($\sim1$
Gy) than the spiral and ocular filaments in IC2163/NGC 2207,
however, so corrections for cluster dissolution had to be applied
in that study.

The distinction between power law and peaked mass distributions
for individual clusters and star complexes may have important
implications for the formation of globular clusters in the early
Universe.  Old globular clusters currently have Gaussian
distribution functions for their magnitudes, unlike young clusters
and super star clusters, which have power laws. One solution is
that the low mass halo globulars have evaporated by now (Okazaki
\& Tosa 1995; Baumgardt 1998), but this implies there should be a
correlation between the peak in the globular cluster luminosity
function and the galactocentric distance (Vesperini 2000a,b),
because clusters closer to the center should evolve faster. Such a
correlation is not observed (Kundu et al. 1999; see review in Fall
\& Zhang 2002).  Parmentier \& Gilmore (2005) recently simulated
an evolving population of clusters and concluded that the halo
population started with a Gaussian mass function rather than a
power law. One explanation for this Gaussian could be that halo
globulars form as a result of an instability with a dominant scale
(e.g., Fall \& Rees 1985). This explanation is supported somewhat
by the present observations, although the geometry and instability
are different here. Nevertheless, it seems reasonable that
different circumstances can give different cluster mass
distribution functions.  Perhaps some halo globular clusters
formed by regular instabilities in the spiral arms of interacting
young disk galaxies.

\section{MIPS emission}

The total IR luminosity ($L_{\rm TIR}$) of a region can be
estimated from MIPS emission using the method of Dale \& Helou
(2002):
\begin{equation} L_{\rm TIR} = 1.559\nu L_{24}+0.7686\nu
L_{70}+1.347\nu L_{160}.\end{equation} To determine the luminosity
for both galaxies combined, we used {\it imstat} to define a box
outlining them and subtracted sky background. The luminosity for
both galaxies is $L_{\rm TIR}$=8.1$\times10^{10}$ L$_\odot$.  The
$L_{\rm TIR}$ for the whole inner region of IC 2163, including the
eyelid, is $2.0\times 10^{10}$ L$_\odot$, or about 25\% of the
total from both galaxies. The $L_{\rm TIR}$ from a region
$24^{\prime\prime}$ in diameter surrounding Feature i is
$3.8\times10^{9}$ L$_\odot$, or 4.5\% of the total. Replacing the
2 saturated pixels in Feature i at 24$\mu$m by the same values as
the surrounding pixels or using double the surface brightness at
the FWHM resulted in an insignificant change in its flux. Feature
i is extraordinary at $24\mu$m, accounting for 12\% of the total
emission in this passband. All of these results use the values in
Table 1.

The total star formation rate can be estimated by multiplying the
infrared luminosity (in cgs units) by a conversion factor of
$4.5\times10^{-44}$, giving $M\odot/$year (Kennicutt 1998). This
conversion was initially based on IRAS results. Detailed UV, IR,
and H$\alpha$ properties of star-forming clumps in M51 by Calzetti
et al. (2005) indicate that MIPS emission (particularly 24$\mu$m)
is a good star formation indicator, but that there are variations
by a factor of several from one galaxy to another because of
different possible sources of MIPS emission.  The result using
$L_{\rm TIR}$ is a star formation rate of $\sim$14.5 M$_\odot$
yr$^{-1}$ for the combined galaxies. For the eyelid, the star
formation rate is $\sim$3.6 M$_\odot$ yr$^{-1}$, and for Feature
i, it is $\sim$0.7 M$_\odot$ yr$^{-1}$. The eyelid and Feature i
contain intense star formation considering their small sizes. In
Paper IV, we found 15 star-forming regions on the HST WFPC2 images
in the vicinity of Feature i included in the SST photometry. Their
sizes ranged from unresolved to a few dozen parsecs. Based on
their absolute V magnitudes and [V-I] colors and a comparison with
Starburst 99 evolution models (Leitherer et al. 1999), the masses
of these individual regions range from $\sim 2\times 10^4$
$M_\odot$ to $\times10^6$ $M_\odot$ with ages of a few million
years.  These values are consistent with the MIPS star formation
rate.

The H$\alpha$ luminosities correlate with $8\mu$m and $24\mu$m
luminosities in star-forming galaxies (Wu et al. 2005). Based on
their conversions, we get a total star formation rate from $8\mu$m
in the combined galaxies equal to $10.8$ M$_\odot$ yr$^{-1}$; from
$24\mu$m, it is $13.7$ M$_\odot$ yr$^{-1}$. These rates are
similar to those calculated above.

\section{Conclusions}

IC 2163 and NGC 2207 contain several hundred clumps of IR emission
strung out along thin spiral or oval-like arcs. One star-forming
region, called Feature i in our previous HST study, dominates the
short-wavelength MIPS images; it accounts for 12\% of the total
24$\mu$m emission from the two galaxies. The next brightest
regions are in the eyelid of IC 2163, where galaxy-scale shocks
formed from the recent interaction with NGC 2207. All of these
regions appear to be young ($\sim 10^7$ yr) based on their optical
and IRAC colors, and their association with H$\alpha$ emission.
The SED for the IC 2163/NGC 2207 pair matches that of other
late-type galaxies observed by SINGS.

The luminosity distribution function of the IRAC clumps in the
eyelid region of IC 2163 shows significant deviations from a power
law. The prevalent clump morphology of beads on a string in this
region and in the spiral arms of NGC 2207 suggests the clumps
formed by regular gravitational instabilities, in which case the
luminosity function is consistent with a characteristic mass from
the instability. This is unlike the usual situation for individual
clusters in galaxy disks where the gas and associated turbulent
motions produce hierarchical structure and power-law mass
distributions. The IRAC clumps are much larger than individual
star clusters. Our peaked distribution for the IRAC clumps should
be compared with the luminosity or mass distributions of giant
star complexes in grand design galaxy spiral arms. The HST cluster
luminosity function in the eyelid region of IC 2163 is in fact
closer to the standard power law, even though the IRAC clumps
there have a peaked distribution with a characteristic luminosity
of $10^8$ L$_\odot$.

We gratefully acknowledge support from NASA JPL/Spitzer grant RSA
1264471 to M.K. and RSA 1264582 to D.M.E. for Cycle 1
observations. We thank Bev Smith for helpful discussions and an
anonymous referee for useful comments. Deidre Hunter kindly
provided the H$\alpha$ image, based on observations at Lowell
Observatory.

\clearpage
\begin{deluxetable}{cccccccc}
\tablecaption{Integrated Region Flux Densities\label{tab:whole}}
\tablehead{ \colhead{Region} &\colhead{S$_{3.6}$}
&\colhead{S$_{4.5}$} &\colhead{S$_{5.8}$} &\colhead{S$_{8.0}$}
&\colhead{S$_{24}$} &\colhead{S$_{70}$} &\colhead{S$_{160}$}\\
&\colhead{mJy} &\colhead{mJy} &\colhead{mJy} &\colhead{mJy}
&\colhead{mJy} &\colhead{mJy} &\colhead{mJy} } \startdata
IC 2163 eyelid\tablenotemark{a}&88.0&59.1&177&476&506&5115&11000\\
Feature i\tablenotemark{b}&2.8&2.5&13.2&38&248&1210&561\\
Combined Galaxies&350&174&1150&1130&2000&24500&40000\\
\enddata
\tablenotetext{a}{Includes eyelid and all of the region interior.}
\tablenotetext{b}{All passbands were measured for a
$24^{\prime\prime}$ box.}
\end{deluxetable}

\clearpage

\begin{deluxetable}{cccccccccc}
\rotate
\tablewidth{0pt}
\tablecaption{Prominent Clump Locations and Flux Densities\label{tab:clump}}
\tablehead{
\colhead{No.}
&\colhead{RA (J2000)}
&\colhead{Dec (J2000)}
&\colhead{S$_{3.6}$}
&\colhead{S$_{4.5}$}
&\colhead{S$_{5.8}$}
&\colhead{S$_{8.0}$}
&\colhead{S$_{24}$\tablenotemark{a}}
&\colhead{S$_{70}$\tablenotemark{a}}
&\colhead{L(H$\alpha$)}\\
&\colhead{$6^{\rm h}16^{\rm m}$}
&\colhead{$-21^\circ$}
&\colhead{mJy}
&\colhead{mJy}
&\colhead{mJy}
&\colhead{mJy}
&\colhead{mJy}
&\colhead{mJy}
&\colhead{$\log L_\odot$}}
\startdata
IC 2163 eyelid&&&&&&&&&\\
    1& $28.907^{\rm s}$&  $22^\prime\;53.54^{\prime\prime}$    &0.80&0.64&3.54&10.36&11.38&-&38.9\\
    2& $27.825^{\rm s}$&  $22^\prime\;50.68^{\prime\prime}$&1.09 &0.78&3.95&11.92&9.89&-&38.5\\
    3& $27.484^{\rm s}$&  $22^\prime\;48.66^{\prime\prime}$&1.05&0.80&4.33&13.48&18.53&-&39.0\\
    4& $27.115^{\rm s}$&  $22^\prime\;43.06^{\prime\prime}$    &0.95&0.73&4.12&12.54&18.82&-&39.6\\
    5& $26.633^{\rm s}$&  $22^\prime\;37.05^{\prime\prime}$  &0.90&0.68&2.99&8.16&16.48&-&39.8\\
    6& $27.062^{\rm s}$&  $22^\prime\;32.31^{\prime\prime}$  &0.48&0.36&1.62&5.47&8.35&-&-\\
    8& $29.001^{\rm s}$&  $22^\prime\;32.05^{\prime\prime}$&1.16&0.81&3.61&10.25&14.67&-&39.1\\
    9& $28.803^{\rm s}$&  $22^\prime\;26.06^{\prime\prime}$&1.72&1.33&6.47&19.34&50.61&333.55&39.6\\
    10& $28.151^{\rm s}$& $22^\prime\;16.86^{\prime\prime}$&1.46&1.13&5.14&15.03&29.80&-&39.3\\
    11& $26.471^{\rm s}$& $22^\prime\;12.76^{\prime\prime}$&1.01&0.77&3.95&11.16&18.72&-&39.5\\
NGC 2207 inner&&&&&&&&&\\
    19& $23.421^{\rm s}$& $22^\prime\;12.54^{\prime\prime}$&1.14&0.82&4.16&11.52&19.05&-&39.8\\
NGC 2207 middle&&&&&&&&&\\
    12& $25.328^{\rm s}$& $22^\prime\;20.24^{\prime\prime}$&0.56&0.42&2.16&6.09&3.00&240.4&39.2\\
NGC 2207 outer&&&&&&&&&\\
    13& $24.883^{\rm s}$& $21^\prime\;50.35^{\prime\prime}$&0.44&0.34&1.76&4.97&11.67&-&39.4\\
    14& $24.484^{\rm s}$& $21^\prime\;49.52^{\prime\prime}$&0.69&0.52&2.61&6.91&-&-&39.6\\
    15& $23.800^{\rm s}$& $21^\prime\;50.27^{\prime\prime}$&0.30&0.28&1.76&4.69&4.80&-&39.1\\
    16& $23.091^{\rm s}$& $21^\prime\;41.86^{\prime\prime}$&0.73&0.58&3.40&9.35&19.32&-&39.6\\
    17& $22.465^{\rm s}$& $21^\prime\;39.02^{\prime\prime}$&0.50&0.42&2.33&6.17&11.64&-&39.8\\
    18& $18.582^{\rm s}$& $21^\prime\;55.85^{\prime\prime}$&0.80&0.64&3.48&9.79&  24.41&-&39.7\\
    20\tablenotemark{b}& $15.900^{\rm s}$& $22^\prime\;2.82^{\prime\prime}$&2.70&2.46&12.16&35.09&277.48&1628.64&40.2\\
\enddata
\tablenotetext{a}{Pixel sizes for $24\mu$m ($2.45^{\prime\prime}$)
and $70\mu$m ($4.0^{\prime\prime}$) emissions are much larger than
for the IRAC bands, which are all $1.2^{\prime\prime}$. A direct
comparison of tabulated IRAC and MIPS flux densities, to give
colors for example, should not be made.} \tablenotetext{b}{Feature
i}
\end{deluxetable}

\clearpage

\begin{deluxetable}{lcccccc}
\tablecaption{Measured Clump Locations and IRAC Flux
Densities\label{tab:clump2}} \tablehead{
\colhead{ID} &
\colhead{RA (J2000)} &
\colhead{Dec (J2000)} &
\colhead{S$_{3.6}$} &
\colhead{S$_{4.5}$} &
\colhead{S$_{5.8}$} &
\colhead{S$_{8.0}$}\\
&\colhead{$6^{\rm h}16^{\rm m}$}
&\colhead{$-21^\circ$}
&\colhead{mJy}
&\colhead{mJy}
&\colhead{mJy}
&\colhead{mJy}}
\startdata
1 & 28.907 &  $22^{\prime}\;\;53.54^{\prime\prime}$ & 0.80 & 0.64 & 3.54 & 10.36 \\
2 & 27.825 &  $22^{\prime}\;\;50.68^{\prime\prime}$ & 1.09 & 0.78 & 3.95 & 11.92 \\
3 & 27.484 &  $22^{\prime}\;\;48.66^{\prime\prime}$ & 1.05 & 0.80 & 4.33 & 13.48 \\
4 & 27.115 &  $22^{\prime}\;\;43.06^{\prime\prime}$ & 0.95 & 0.73 & 4.12 & 12.54 \\
5 & 26.633 &  $22^{\prime}\;\;37.05^{\prime\prime}$ & 0.90 & 0.68 & 2.99 & 8.16 \\
6 & 27.062 &  $22^{\prime}\;\;32.31^{\prime\prime}$ & 0.48 & 0.36 & 1.62 & 5.47 \\
7 & 26.410 &  $22^{\prime}\;\;22.31^{\prime\prime}$ & 0.33 & 0.21 &  - &  - \\
8 & 29.001 &  $22^{\prime}\;\;32.05^{\prime\prime}$ & 1.16 & 0.81 & 3.61 & 10.25 \\
9 & 28.803 &  $22^{\prime}\;\;26.06^{\prime\prime}$ & 1.72 & 1.33 & 6.47 & 19.34 \\
10 & 28.151 &  $22^{\prime}\;\;16.86^{\prime\prime}$ & 1.46 & 1.13 & 5.14 & 15.03 \\
11 & 26.471 &  $22^{\prime}\;\;12.76^{\prime\prime}$ & 1.01 & 0.77 & 3.95 & 11.16 \\
12 & 16.496 &  $22^{\prime}\;\;20.24^{\prime\prime}$ & 0.56 & 0.42 & 2.16 & 6.09 \\
13 & 24.883 &  $21^{\prime}\;\;50.35^{\prime\prime}$ & 0.44 & 0.34 & 1.76 & 4.97 \\
14 & 24.484 &  $21^{\prime}\;\;49.52^{\prime\prime}$ & 0.69 & 0.52 & 2.61 & 6.91 \\
15 & 23.800 &  $21^{\prime}\;\;50.27^{\prime\prime}$ & 0.30 & 0.28 & 1.76 & 4.69 \\
16 & 23.091 &  $21^{\prime}\;\;41.86^{\prime\prime}$ & 0.73 & 0.58 & 3.40 & 9.35 \\
17 & 22.465 &  $21^{\prime}\;\;39.02^{\prime\prime}$ & 0.50 & 0.42 & 2.33 & 6.17 \\
18 & 18.582 &  $21^{\prime}\;\;55.85^{\prime\prime}$ & 0.80 & 0.64 & 3.48 & 9.79 \\
19 & 23.421 &  $22^{\prime}\;\;12.54^{\prime\prime}$ & 1.14 & 0.82 & 4.16 & 11.52 \\
20\tablenotemark{a} & 15.900 &  $22^{\prime}\;\;2.82^{\prime\prime}$ & 2.70 & 2.46 & 12.16 & 35.09 \\
21 & 16.096 &  $22^{\prime}\;\;10.39^{\prime\prime}$ & 0.54 & 0.39 & 1.86 & 5.25 \\
22 & 16.496 &  $22^{\prime}\;\;9.23^{\prime\prime}$ & 0.44 & 0.32 & 1.54 & 4.51 \\
23 & 16.320 &  $22^{\prime}\;\;21.96^{\prime\prime}$ & 0.22 & 0.17 & 1.04 & 2.93 \\
24 & 16.631 &  $22^{\prime}\;\;29.15^{\prime\prime}$ & 0.29 & 0.24 & 1.25 & 3.56 \\
25 & 17.139 &  $22^{\prime}\;\;42.72^{\prime\prime}$ & 0.27 & 0.21 & 1.19 & 3.09 \\
26 & 17.594 &  $22^{\prime}\;\;44.35^{\prime\prime}$ & 0.54 & 0.38 & 2.08 & 5.65 \\
27 & 18.340 &  $22^{\prime}\;\;32.46^{\prime\prime}$ & 0.38 & 0.28 & 1.80 & 5.05 \\
28 & 18.709 &  $22^{\prime}\;\;35.67^{\prime\prime}$ & 0.42 & 0.31 & 1.81 & 5.05 \\
29 & 18.709 &  $22^{\prime}\;\;36.10^{\prime\prime}$ & 0.48 & 0.33 & 1.63 & 4.93 \\
30 & 19.619 &  $22^{\prime}\;\;40.12^{\prime\prime}$ & 0.43 & 0.37 & 2.02 & 5.91 \\
31 & 20.273 &  $22^{\prime}\;\;46.54^{\prime\prime}$ & 0.29 & 0.21 & 1.13 & 3.21 \\
32 & 21.529 &  $22^{\prime}\;\;42.26^{\prime\prime}$ & 0.46 & 0.29 & 1.42 & 4.76 \\
33 & 23.215 &  $22^{\prime}\;\;29.64^{\prime\prime}$ & 0.31 & 0.20 & 1.42 & 4.39 \\
34 & 24.025 &  $21^{\prime}\;\;59.44^{\prime\prime}$ & 0.35 & 0.27 & 1.46 & 3.59 \\
35 & 23.683 &  $21^{\prime}\;\;58.62^{\prime\prime}$ & 0.23 & 0.17 & 0.62 & 1.32 \\
36 & 23.341 &  $21^{\prime}\;\;58.20^{\prime\prime}$ & 0.12 & 0.10 & 0.77 & 1.53 \\
37 & 25.253 &  $21^{\prime}\;\;52.76^{\prime\prime}$ & 0.31 & 0.22 & 1.33 & 3.67 \\
38 & 24.002 &  $21^{\prime}\;\;43.12^{\prime\prime}$ & 0.42 & 0.32 & 1.27 & 3.50 \\
39 & 21.694 &  $21^{\prime}\;\;40.56^{\prime\prime}$ & 0.11 & 0.07 & 0.57 & 1.79 \\
40 & 21.038 &  $21^{\prime}\;\;43.30^{\prime\prime}$ & 0.14 & 0.09 & 0.78 & 2.36 \\
41 & 20.639 &  $21^{\prime}\;\;44.06^{\prime\prime}$ & 0.15 & 0.11 & 0.82 & 2.29 \\
42 & 20.382 &  $21^{\prime}\;\;44.44^{\prime\prime}$ & 0.06 & 0.04 & 0.41 & 1.09 \\
43 & 19.526 &  $21^{\prime}\;\;46.37^{\prime\prime}$ & 0.22 & 0.15 & 0.78 & 2.39 \\
44 & 19.211 &  $21^{\prime}\;\;49.13^{\prime\prime}$ & 0.32 & 0.22 & 1.16 & 3.32 \\
45 & 19.211 &  $21^{\prime}\;\;51.10^{\prime\prime}$ & 0.40 & 0.31 & 1.70 & 4.84 \\
46 & 19.151 &  $21^{\prime}\;\;59.08^{\prime\prime}$ & 0.14 & 0.10 & 0.57 & 1.95 \\
47 & 18.607 &  $22^{\prime}\;\;03.42^{\prime\prime}$ & 0.19 & 0.13 & 0.85 & 2.26 \\
48 & 17.865 &  $22^{\prime}\;\;06.15^{\prime\prime}$ & 0.30 & 0.22 & 0.58 & 1.78 \\
49 & 17.607 &  $22^{\prime}\;\;10.91^{\prime\prime}$ & 0.26 & 0.20 & 1.01 & 2.63 \\
50 & 18.118 &  $22^{\prime}\;\;14.93^{\prime\prime}$ & 0.24 & 0.18 & 0.82 & 2.56 \\
51 & 18.544 &  $22^{\prime}\;\;21.33^{\prime\prime}$ & 0.21 & 0.15 & 0.63 & 1.60 \\
52 & 21.047 &  $22^{\prime}\;\;35.05^{\prime\prime}$ & 0.42 & 0.33 & 0.99 & 2.66 \\
53 & 21.872 &  $22^{\prime}\;\;39.49^{\prime\prime}$ & 0.42 & 0.26 & 0.75 & 2.68 \\
54 & 23.831 &  $22^{\prime}\;\;59.54^{\prime\prime}$ & 0.26 & 0.27 & 1.12 & 2.88 \\
55 & 23.571 &  $23^{\prime}\;\;9.88^{\prime\prime}$ & 0.20 & 0.14 & 0.57 & 1.62 \\
56 & 24.112 &  $23^{\prime}\;\;9.52^{\prime\prime}$ & 0.22 & 0.15 & 0.75 & 2.08 \\
57 & 25.208 &  $22^{\prime}\;\;36.95^{\prime\prime}$ & 0.33 & 0.23 & 1.14 & 3.04 \\
58 & 25.182 &  $22^{\prime}\;\;29.39^{\prime\prime}$ & 0.40 & 0.31 & 1.59 & 4.31 \\
59 & 29.363 &  $22^{\prime}\;\;54.77^{\prime\prime}$ & 0.68 & 0.50 & 2.28 & 6.67 \\
60 & 30.876 &  $22^{\prime}\;\;48.90^{\prime\prime}$ & 0.23 & 0.16 & 0.76 & 2.08 \\
61 & 31.418 &  $22^{\prime}\;\;47.75^{\prime\prime}$ & 0.36 & 0.24 & 0.97 & 2.93 \\
62 & 23.844 &  $22^{\prime}\;\;24.91^{\prime\prime}$ & 0.21 & 0.14 & 0.75 & 2.01 \\
63 & 23.501 &  $22^{\prime}\;\;26.08^{\prime\prime}$ & 0.11 & 0.08 & 0.43 & 1.07 \\
64 & 30.935 &  $22^{\prime}\;\;42.14^{\prime\prime}$ & 0.03 & 0.04 & 0.27 & 0.53 \\
65 & 30.625 &  $22^{\prime}\;\;32.56^{\prime\prime}$ & 0.00 & 0.00 & 0.02 & 0.07 \\
66 & 30.478 &  $22^{\prime}\;\;45.69^{\prime\prime}$ & 0.00 & 0.00 & 0.06 & 0.32 \\
67 & 29.624 &  $22^{\prime}\;\;42.44^{\prime\prime}$ & 0.04 & 0.02 & 0.00 & 0.00 \\
68 & 29.854 &  $22^{\prime}\;\;36.89^{\prime\prime}$ & 0.04 & 0.08 & 0.51 & 1.38 \\
69 & 31.875 &  $22^{\prime}\;\;45.39^{\prime\prime}$ & 0.21 & 0.12 & 0.67 & 2.14 \\
70 & 32.417 &  $22^{\prime}\;\;43.44^{\prime\prime}$ & 0.20 & 0.13 & 0.62 & 1.89 \\
71 & 28.998 &  $22^{\prime}\;\;38.42^{\prime\prime}$ & 0.00 & 0.00 & 0.00 & 0.00 \\
72 & 29.369 &  $22^{\prime}\;\;36.85^{\prime\prime}$ & 0.25 & 0.15 & 0.45 & 1.48 \\
73 & 28.824 &  $22^{\prime}\;\;47.56^{\prime\prime}$ & 0.00 & 0.00 & 0.11 & 0.52 \\
74 & 28.397 &  $22^{\prime}\;\;44.75^{\prime\prime}$ & 0.00 & 0.00 & 0.35 & 1.60 \\
75 & 28.111 &  $22^{\prime}\;\;47.91^{\prime\prime}$ & 0.11 & 0.02 & 0.27 & 1.58 \\
76 & 28.884 &  $22^{\prime}\;\;40.40^{\prime\prime}$ & 0.24 & 0.15 & 0.01 & 0.67 \\
77 & 27.459 &  $22^{\prime}\;\;37.11^{\prime\prime}$ & 0.00 & 0.00 & 0.00 & 0.00 \\
78 & 27.032 &  $22^{\prime}\;\;37.88^{\prime\prime}$ & 1.13 & 0.76 & 2.25 & 7.89 \\
79 & 27.206 &  $22^{\prime}\;\;27.54^{\prime\prime}$ & 0.93 & 0.54 & 1.69 & 5.67 \\
80 & 27.892 &  $22^{\prime}\;\;22.81^{\prime\prime}$ & 0.96 & 0.67 & 3.67 & 10.93 \\
81 & 27.069 &  $22^{\prime}\;\;13.99^{\prime\prime}$ & 0.82 & 0.60 & 3.48 & 9.59 \\
82 & 31.682 &  $22^{\prime}\;\;25.47^{\prime\prime}$ & 0.06 & 0.02 & 0.02 & 0.05 \\
83 & 31.197 &  $22^{\prime}\;\;29.02^{\prime\prime}$ & 0.07 & 0.05 & 0.10 & 0.29 \\
84 & 30.829 &  $22^{\prime}\;\;21.03^{\prime\prime}$ & 0.23 & 0.18 & 1.20 & 3.19 \\
85 & 33.418 &  $22^{\prime}\;\;34.35^{\prime\prime}$ & 0.06 & 0.04 & 0.16 & 0.51 \\
86 & 33.934 &  $22^{\prime}\;\;25.63^{\prime\prime}$ & 0.13 & 0.09 & 0.50 & 1.26 \\
87 & 33.451 &  $22^{\prime}\;\;21.61^{\prime\prime}$ & 0.03 & 0.03 & 0.03 & 0.22 \\
88 & 33.108 &  $22^{\prime}\;\;34.38^{\prime\prime}$ & 0.06 & 0.05 & 0.27 & 0.79 \\
89 & 23.489 &  $21^{\prime}\;\;43.08^{\prime\prime}$ & 0.54 & 0.41 & 2.03 & 6.01 \\
90 & 22.720 &  $21^{\prime}\;\;40.63^{\prime\prime}$ & 0.53 & 0.43 & 2.34 & 6.32 \\
91 & 18.241 &  $21^{\prime}\;\;53.84^{\prime\prime}$ & 0.24 & 0.20 & 1.07 & 2.82 \\
92 & 17.776 &  $22^{\prime}\;\;16.89^{\prime\prime}$ & 0.16 & 0.12 & 0.80 & 2.23 \\
93 & 20.311 &  $22^{\prime}\;\;20.27^{\prime\prime}$ & 0.49 & 0.31 & 0.98 & 2.59 \\
94 & 20.572 &  $22^{\prime}\;\;09.14^{\prime\prime}$ & 0.23 & 0.14 & 0.75 & 2.37 \\
95 & 21.002 &  $22^{\prime}\;\;02.80^{\prime\prime}$ & 0.09 & 0.07 & 0.28 & 0.88 \\
96 & 21.859 &  $21^{\prime}\;\;58.49^{\prime\prime}$ & 0.13 & 0.08 & 0.27 & 0.82 \\
97 & 22.344 &  $21^{\prime}\;\;57.73^{\prime\prime}$ & 0.08 & 0.06 & 0.21 & 0.74 \\
98 & 22.431 &  $21^{\prime}\;\;53.75^{\prime\prime}$ & 0.08 & 0.06 & 0.17 & 0.63 \\
99 & 22.686 &  $21^{\prime}\;\;55.36^{\prime\prime}$ & 0.11 & 0.07 & 0.39 & 1.17 \\
100 & 22.830 &  $21^{\prime}\;\;53.38^{\prime\prime}$ & 0.06 & 0.04 & 0.23 & 0.97 \\
101 & 24.899 &  $22^{\prime}\;\;23.39^{\prime\prime}$ & 0.35 & 0.26 & 1.10 & 2.95 \\
102 & 24.929 &  $22^{\prime}\;\;18.62^{\prime\prime}$ & 0.56 & 0.41 & 1.89 & 5.37 \\
103 & 24.731 &  $22^{\prime}\;\;14.62^{\prime\prime}$ & 0.20 & 0.16 & 0.84 & 2.35 \\
104 & 24.932 &  $22^{\prime}\;\;12.65^{\prime\prime}$ & 0.59 & 0.40 & 2.12 & 6.30 \\
105 & 24.620 &  $22^{\prime}\;\;06.65^{\prime\prime}$ & 0.33 & 0.24 & 1.87 & 4.93 \\
106 & 25.892 &  $22^{\prime}\;\;37.40^{\prime\prime}$ & 0.10 & 0.05 & 0.74 & 2.07 \\
107 & 24.889 &  $22^{\prime}\;\;49.67^{\prime\prime}$ & 0.07 & 0.05 & 0.22 & 0.66 \\
108 & 25.827 &  $22^{\prime}\;\;57.70^{\prime\prime}$ & 0.12 & 0.08 & 0.48 & 1.38 \\
109 & 25.791 &  $23^{\prime}\;\;16.81^{\prime\prime}$ & 0.06 & 0.04 & 0.13 & 0.29 \\
100 & 23.036 &  $22^{\prime}\;\;50.73^{\prime\prime}$ & 0.05 & 0.03 & 0.15 & 0.32 \\
111 & 22.471 &  $22^{\prime}\;\;38.34^{\prime\prime}$ & 0.27 & 0.20 & 0.90 & 2.89 \\
112 & 23.476 &  $22^{\prime}\;\;18.91^{\prime\prime}$ & 0.08 & 0.04 & 0.68 & 1.83 \\
113 & 18.484 &  $22^{\prime}\;\;27.70^{\prime\prime}$ & 0.17 & 0.10 & 0.17 & 0.77 \\
114 & 21.432 &  $23^{\prime}\;\;11.71^{\prime\prime}$ & 0.51 & 0.40 & 2.08 & 5.69 \\
115 & 19.242 &  $22^{\prime}\;\;57.61^{\prime\prime}$ & 0.23 & 0.20 & 0.78 & 1.98 \\
116 & 19.015 &  $22^{\prime}\;\;55.21^{\prime\prime}$ & 0.16 & 0.11 & 0.66 & 1.70 \\
117 & 17.477 &  $22^{\prime}\;\;51.51^{\prime\prime}$ & 0.25 & 0.19 & 0.80 & 2.18 \\
118 & 17.647 &  $22^{\prime}\;\;55.10^{\prime\prime}$ & 0.16 & 0.13 & 0.53 & 1.44 \\
119 & 17.249 &  $22^{\prime}\;\;50.69^{\prime\prime}$ & 0.14 & 0.10 & 0.60 & 1.46 \\
120 & 18.013 &  $21^{\prime}\;\;52.62^{\prime\prime}$ & 0.29 & 0.23 & 1.22 & 3.43 \\
121 & 15.704 &  $21^{\prime}\;\;53.25^{\prime\prime}$ & 0.19 & 0.15 & 0.90 & 2.60 \\
122 & 15.649 &  $21^{\prime}\;\;48.07^{\prime\prime}$ & 0.10 & 0.09 & 0.54 & 1.61 \\
123 & 16.742 &  $22^{\prime}\;\;35.92^{\prime\prime}$ & 0.18 & 0.13 & 0.83 & 2.14 \\
124 & 16.122 &  $22^{\prime}\;\;17.17^{\prime\prime}$ & 0.17 & 0.12 & 0.78 & 1.95 \\
125 & 20.578 &  $23^{\prime}\;\;09.26^{\prime\prime}$ & 0.13 & 0.10 & 0.49 & 1.31 \\
126 & 22.285 &  $23^{\prime}\;\;17.74^{\prime\prime}$ & 0.11 & 0.08 & 0.29 & 0.79 \\
127 & 26.817 &  $23^{\prime}\;\;18.47^{\prime\prime}$ & 0.08 & 0.05 & 0.25 & 0.68 \\
128 & 24.251 &  $23^{\prime}\;\;19.88^{\prime\prime}$ & 0.09 & 0.06 & 0.00 & 0.00 \\
129 & 24.649 &  $23^{\prime}\;\;23.49^{\prime\prime}$ & 0.08 & 0.05 & 0.24 & 0.64 \\
130 & 24.703 &  $23^{\prime}\;\;31.46^{\prime\prime}$ & 0.10 & 0.08 & 0.26 & 0.91 \\
131 & 25.044 &  $23^{\prime}\;\;34.27^{\prime\prime}$ & 0.16 & 0.10 & 0.03 & 0.03 \\
132 & 21.579 &  $23^{\prime}\;\;00.57^{\prime\prime}$ & 0.20 & 0.16 & 0.26 & 0.62 \\
133 & 23.926 &  $22^{\prime}\;\;34.07^{\prime\prime}$ & 0.10 & 0.08 & 0.14 & 0.56 \\
134 & 28.593 &  $22^{\prime}\;\;54.32^{\prime\prime}$ & 0.06 & 0.12 & 1.19 & 2.93 \\
135 & 28.138 &  $22^{\prime}\;\;52.89^{\prime\prime}$ & 0.57 & 0.47 & 2.19 & 5.83 \\
136 & 25.65  &  $21^{\prime}\;\;56.37^{\prime\prime}$ & 0.18 & 0.15 & 0.62 & 1.67 \\
137 & 16.073 &  $21^{\prime}\;\;58.45^{\prime\prime}$ & 0.65 & 0.40 & 0.62 & 3.29 \\
138 & 17.136 &  $21^{\prime}\;\;36.63^{\prime\prime}$ & 0.19 & 0.16 & 0.39 & 1.04 \\
139 & 17.474 &  $21^{\prime}\;\;47.41^{\prime\prime}$ & 0.13 & 0.11 & 0.04 & 0.00 \\
140 & 17.618 &  $21^{\prime}\;\;42.64^{\prime\prime}$ & 0.12 & 0.08 & 0.40 & 1.05 \\
141 & 16.719 &  $22^{\prime}\;\;21.19^{\prime\prime}$ & 0.14 & 0.10 & 0.40 & 1.03 \\
142 & 25.501 &  $22^{\prime}\;\;13.48^{\prime\prime}$ & 0.44 & 0.33 & 0.99 & 0.00 \\
143 & 25.240 &  $22^{\prime}\;\;25.41^{\prime\prime}$ & 0.21 & 0.13 & 0.50 & 1.83 \\
144 & 29.003 &  $22^{\prime}\;\;24.49^{\prime\prime}$ & 1.09 & 0.79 & 3.04 & 10.04 \\
145 & 28.371 &  $22^{\prime}\;\;38.18^{\prime\prime}$ & 0.47 & 0.28 & 1.05 & 1.32 \\
146 & 25.727 &  $22^{\prime}\;\;18.87^{\prime\prime}$ & 0.13 & 0.10 & 0.30 & 1.02 \\
147 & 15.963 &  $22^{\prime}\;\;23.32^{\prime\prime}$ & 0.13 & 0.08 & 0.37 & 1.22 \\
148 & 16.132 &  $22^{\prime}\;\;28.71^{\prime\prime}$ & 0.10 & 0.12 & 0.31 & 0.84 \\
149 & 16.016 &  $22^{\prime}\;\;33.08^{\prime\prime}$ & 0.06 & 0.03 & 0.09 & 0.28 \\
150 & 16.314 &  $22^{\prime}\;\;36.29^{\prime\prime}$ & 0.04 & 0.04 & 0.06 & 0.00 \\
151 & 20.720 &  $22^{\prime}\;\;32.04^{\prime\prime}$ & 0.22 & 0.16 & 0.42 & 1.78 \\
152 & 20.420 &  $22^{\prime}\;\;33.41^{\prime\prime}$ & 0.71 & 0.48 & 1.30 & 5.24 \\
153 & 21.100 &  $22^{\prime}\;\;43.82^{\prime\prime}$ & 0.11 & 0.08 & 0.58 & 1.59 \\
154 & 20.744 &  $22^{\prime}\;\;42.79^{\prime\prime}$ & 0.07 & 0.05 & 0.39 & 1.49 \\
155 & 21.099 &  $21^{\prime}\;\;31.56^{\prime\prime}$ & 0.11 & 0.06 & 0.02 & 0.00 \\
156 & 20.065 &  $21^{\prime}\;\;52.58^{\prime\prime}$ & 0.08 & 0.04 & 0.02 & 0.00 \\
157 & 20.608 &  $21^{\prime}\;\;49.43^{\prime\prime}$ & 0.04 & 0.03 & 0.06 & 0.23 \\
158 & 22.106 &  $21^{\prime}\;\;45.56^{\prime\prime}$ & 0.05 & 0.04 & 0.05 & 0.32 \\
159 & 21.462 &  $22^{\prime}\;\;29.71^{\prime\prime}$ & 0.38 & 0.19 & 0.00 & 0.00 \\
160 & 33.661 &  $22^{\prime}\;\;31.98^{\prime\prime}$ & 0.05 & 0.05 & 0.21 & 0.57 \\
161 & 34.047 &  $22^{\prime}\;\;30.21^{\prime\prime}$ & 0.11 & 0.07 & 0.16 & 0.37 \\
162 & 23.070 &  $23^{\prime}\;\;14.62^{\prime\prime}$ & 0.10 & 0.06 & 0.00 & 0.23 \\
163 & 22.656 &  $23^{\prime}\;\;16.58^{\prime\prime}$ & 0.09 & 0.06 & 0.15 & 0.64 \\
164 & 22.911 &  $23^{\prime}\;\;21.37^{\prime\prime}$ & 0.10 & 0.06 & 0.21 & 0.57 \\
165 & 22.766 &  $23^{\prime}\;\;26.54^{\prime\prime}$ & 0.10 & 0.08 & 0.06 & 0.06 \\
166 & 26.354 &  $22^{\prime}\;\;59.73^{\prime\prime}$ & 0.09 & 0.07 & 0.30 & 0.79 \\
167 & 26.566 &  $23^{\prime}\;\;4.52^{\prime\prime}$ & 0.02 & 0.00 & 0.05 & 0.28 \\
168 & 26.278 &  $23^{\prime}\;\;12.26^{\prime\prime}$ & 0.07 & 0.05 & 0.00 & 0.00 \\
169 & 26.412 &  $22^{\prime}\;\;55.35^{\prime\prime}$ & 0.03 & 0.03 & 0.22 & 0.50 \\
170 & 26.611 &  $22^{\prime}\;\;57.56^{\prime\prime}$ & 0.00 & 0.00 & 0.00 & 0.00 \\
171 & 26.622 &  $23^{\prime}\;\;6.91^{\prime\prime}$ & 0.05 & 0.02 & 0.23 & 0.54 \\
172 & 26.893 &  $23^{\prime}\;\;5.74^{\prime\prime}$ & 0.00 & 0.00 & 0.15 & 0.31 \\
173 & 27.136 &  $23^{\prime}\;\;3.57^{\prime\prime}$ & 0.02 & 0.02 & 0.07 & 0.33 \\
174 & 26.672 &  $23^{\prime}\;\;24.83^{\prime\prime}$ & 0.03 & 0.02 & 0.08 & 0.15 \\
175 & 26.360 &  $23^{\prime}\;\;21.23^{\prime\prime}$ & 0.01 & 0.01 & 0.05 & 0.06 \\
176 & 25.716 &  $23^{\prime}\;\;26.75^{\prime\prime}$ & 0.06 & 0.04 & 0.12 & 0.51 \\
177 & 25.743 &  $23^{\prime}\;\;32.13^{\prime\prime}$ & 0.03 & 0.01 & 0.09 & 0.19 \\
178 & 25.374 &  $23^{\prime}\;\;27.53^{\prime\prime}$ & 0.06 & 0.06 & 0.08 & 0.07 \\
179 & 25.668 &  $23^{\prime}\;\;40.88^{\prime\prime}$ & 0.05 & 0.03 & 0.10 & 0.41 \\
180 & 25.609 &  $23^{\prime}\;\;47.05^{\prime\prime}$ & 0.04 & 0.02 & 0.00 & 0.00 \\
181 & 24.330 &  $23^{\prime}\;\;36.21^{\prime\prime}$ & 0.05 & 0.04 & 0.02 & 0.00 \\
182 & 23.469 &  $23^{\prime}\;\;14.05^{\prime\prime}$ & 0.14 & 0.09 & 0.23 & 0.78 \\
183 & 23.425 &  $23^{\prime}\;\;16.83^{\prime\prime}$ & 0.11 & 0.06 & 0.26 & 0.73 \\
184 & 23.567 &  $23^{\prime}\;\;20.03^{\prime\prime}$ & 0.17 & 0.12 & 0.53 & 1.39 \\
185 & 24.400 &  $23^{\prime}\;\;02.77^{\prime\prime}$ & 0.06 & 0.04 & 0.10 & 0.29 \\
186 & 22.723 &  $23^{\prime}\;\;26.34^{\prime\prime}$ & 0.11 & 0.09 & 0.07 & 0.10 \\
187 & 24.019 &  $23^{\prime}\;\;31.41^{\prime\prime}$ & 0.00 & 0.00 & 0.00 & 0.00 \\
188 & 21.135 &  $23^{\prime}\;\;03.92^{\prime\prime}$ & 0.08 & 0.06 & 0.28 & 0.61 \\
189 & 24.044 &  $23^{\prime}\;\;01.55^{\prime\prime}$ & 0.22 & 0.15 & 0.62 & 1.77 \\
190 & 23.933 &  $22^{\prime}\;\;54.18^{\prime\prime}$ & 0.13 & 0.08 & 0.39 & 1.08 \\
191 & 24.274 &  $22^{\prime}\;\;55.79^{\prime\prime}$ & 0.07 & 0.07 & 0.31 & 0.75 \\
192 & 25.846 &  $22^{\prime}\;\;45.36^{\prime\prime}$ & 0.02 & 0.03 & 0.03 & 0.00 \\
193 & 25.390 &  $22^{\prime}\;\;44.73^{\prime\prime}$ & 0.07 & 0.04 & 0.11 & 0.02 \\
194 & 25.413 &  $23^{\prime}\;\;0.26^{\prime\prime}$ & 0.05 & 0.05 & 0.18 & 0.56 \\
195 & 25.295 &  $23^{\prime}\;\;9.41^{\prime\prime}$ & 0.06 & 0.06 & 0.11 & 0.09 \\
196 & 21.711 &  $23^{\prime}\;\;26.46^{\prime\prime}$ & 0.13 & 0.06 & 0.07 & 0.28 \\
197 & 23.236 &  $22^{\prime}\;\;48.95^{\prime\prime}$ & 0.05 & 0.04 & 0.09 & 0.28 \\
198 & 23.922 &  $22^{\prime}\;\;44.22^{\prime\prime}$ & 0.05 & 0.04 & 0.09 & 1.12 \\
199 & 26.790 &  $23^{\prime}\;\;13.69^{\prime\prime}$ & 0.12 & 0.09 & 0.34 & 0.64 \\
200 & 19.993 &  $23^{\prime}\;\;9.01^{\prime\prime}$ & 0.06 & 0.05 & 0.22 & 0.62 \\
201 & 19.868 &  $23^{\prime}\;\;0.84^{\prime\prime}$ & 0.07 & 0.06 & 0.21 & 0.41 \\
202 & 18.496 &  $23^{\prime}\;\;11.09^{\prime\prime}$ & 0.08 & 0.06 & 0.15 & 0.00 \\
203 & 19.468 &  $23^{\prime}\;\; 3.80^{\prime\prime}$ & 0.08 & 0.05 & 0.00 & 0.05 \\
204 & 19.183 &  $23^{\prime}\;\; 2.98^{\prime\prime}$ & 0.14 & 0.10 & 0.04 & 0.30 \\
205 & 18.531 &  $22^{\prime}\;\;52.58^{\prime\prime}$ & 0.09 & 0.07 & 0.18 & 0.37 \\
206 & 17.107 &  $22^{\prime}\;\;13.66^{\prime\prime}$ & 0.06 & 0.03 & 0.15 & 1.62 \\
207 & 16.766 &  $22^{\prime}\;\;11.44^{\prime\prime}$ & 0.12 & 0.08 & 0.44 & 0.00 \\
208 & 16.882 &  $22^{\prime}\;\;07.27^{\prime\prime}$ & 0.11 & 0.06 & 0.00 & 0.00 \\
209 & 16.643 &  $21^{\prime}\;\;57.90^{\prime\prime}$ & 0.18 & 0.10 & 0.03 & 0.00 \\
210 & 16.932 &  $21^{\prime}\;\;46.77^{\prime\prime}$ & 0.09 & 0.07 & 0.31 & 0.88 \\
211 & 17.858 &  $21^{\prime}\;\;48.03^{\prime\prime}$ & 0.16 & 0.12 & 0.46 & 1.35 \\
212 & 17.109 &  $21^{\prime}\;\;31.06^{\prime\prime}$ & 0.14 & 0.12 & 0.21 & 0.59 \\
213 & 18.320 &  $21^{\prime}\;\;32.34^{\prime\prime}$ & 0.21 & 0.12 & 0.04 & 0.04 \\
214 & 18.633 &  $21^{\prime}\;\;34.76^{\prime\prime}$ & 0.05 & 0.05 & 0.21 & 0.55 \\
215 & 18.947 &  $21^{\prime}\;\;33.39^{\prime\prime}$ & 0.04 & 0.04 & 0.15 & 0.26 \\
216 & 19.173 &  $21^{\prime}\;\;36.79^{\prime\prime}$ & 0.07 & 0.06 & 0.23 & 0.61 \\
217 & 20.042 &  $21^{\prime}\;\;38.44^{\prime\prime}$ & 0.17 & 0.12 & 0.06 & 0.00 \\
218 & 19.550 &  $21^{\prime}\;\;21.09^{\prime\prime}$ & 0.06 & 0.06 & 0.05 & 0.06 \\
219 & 20.561 &  $21^{\prime}\;\;22.16^{\prime\prime}$ & 0.09 & 0.06 & 0.20 & 0.64 \\
220 & 23.187 &  $21^{\prime}\;\;50.22^{\prime\prime}$ & 0.08 & 0.05 & 0.29 & 1.03 \\
221 & 21.508 &  $21^{\prime}\;\;43.13^{\prime\prime}$ & 0.12 & 0.09 & 0.53 & 1.25 \\
222 & 23.652 &  $22^{\prime}\;\;05.58^{\prime\prime}$ & 0.24 & 0.19 & 1.12 & 2.84 \\
223 & 26.410 &  $22^{\prime}\;\;22.31^{\prime\prime}$ & 1.38 & 0.82 & 0.25 & 0.03 \\
224 & 31.657 &  $22^{\prime}\;\;15.91^{\prime\prime}$ & 0.05 & 0.04 & 0.00 & 0.00 \\
225 & 32.427 &  $22^{\prime}\;\;16.96^{\prime\prime}$ & 0.05 & 0.03 & 0.02 & 0.01 \\

\enddata
\tablenotetext{a}{Feature i}
\end{deluxetable}

\clearpage

\begin{figure}
\epsscale{1.0} \plotone{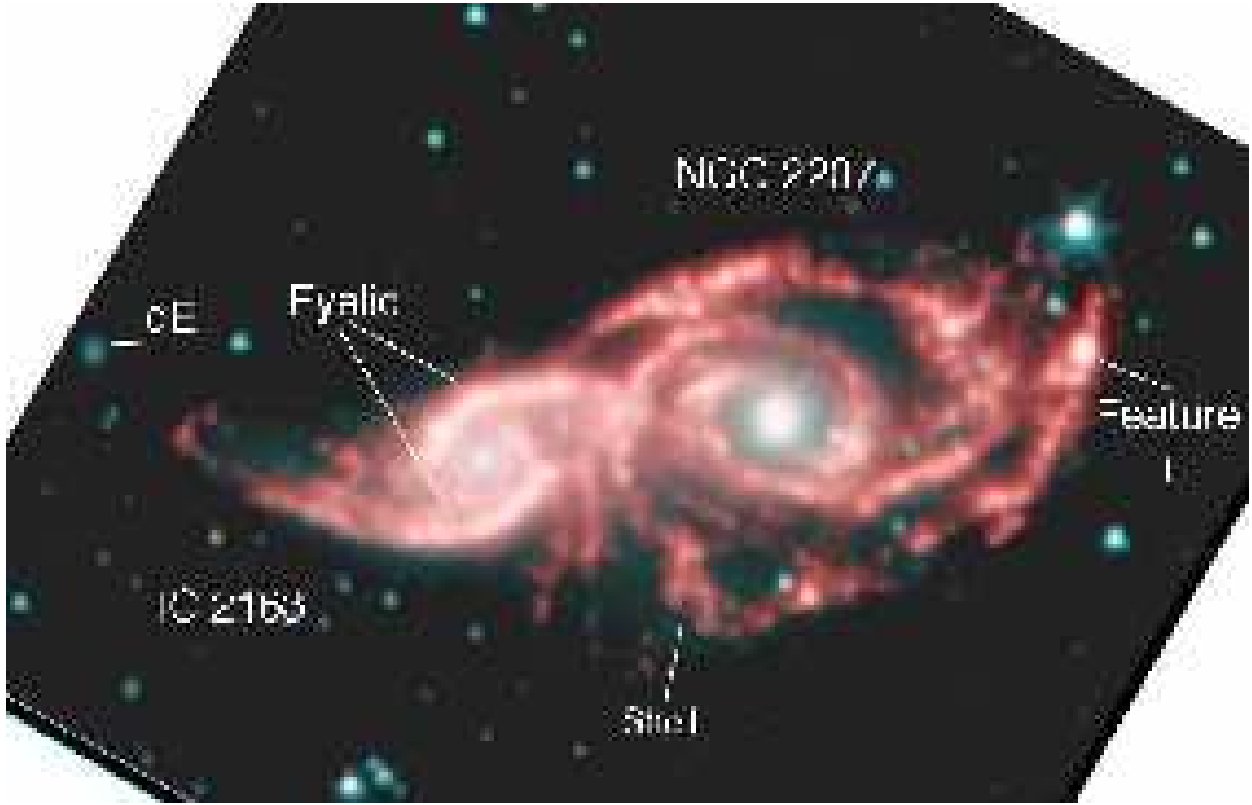} \caption{A composite color image
of the four IRAC images is shown, with bluer colors for shorter
wavelengths and redder for longer. The dwarf galaxy in the east
has an IRAC luminosity of $\sim10^6$ $L_\odot$, assuming it is at
the same distance of 35~Mpc. Its [3.6]-[4.5] color of -0.17 mag is
the same as that observed in a sample of galaxies of morphological
T-type 5 (Pahre et al. 2004b), consistent with it being an
elliptical. (Image degraded for astroph)} \label{fig:color}\end{figure}

\begin{figure}
\epsscale{.8} \plotone{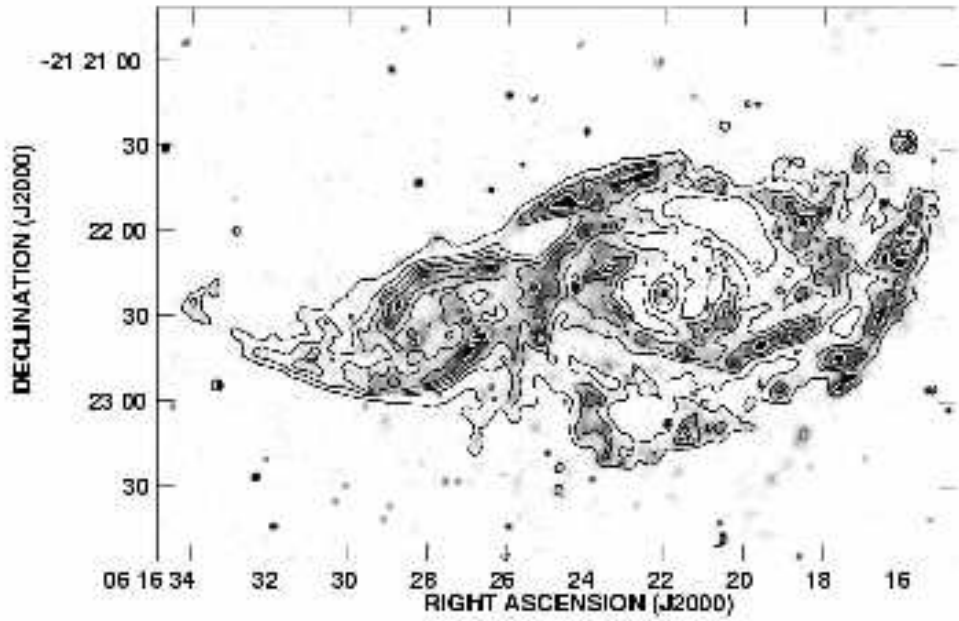} \caption{The continuum-subtracted
H$\alpha$ grayscale is shown with contours from the 8$\mu$m
emission to emphasize the good correspondence of the peaks.
Contour levels for $8\mu$m are 0.5, 2, 4, 8, 16, 32, and 64
MJy/sr. (Image degraded for astroph)} \label{fig:hairac4}\end{figure}

\begin{figure}
\epsscale{1.0} \plotone{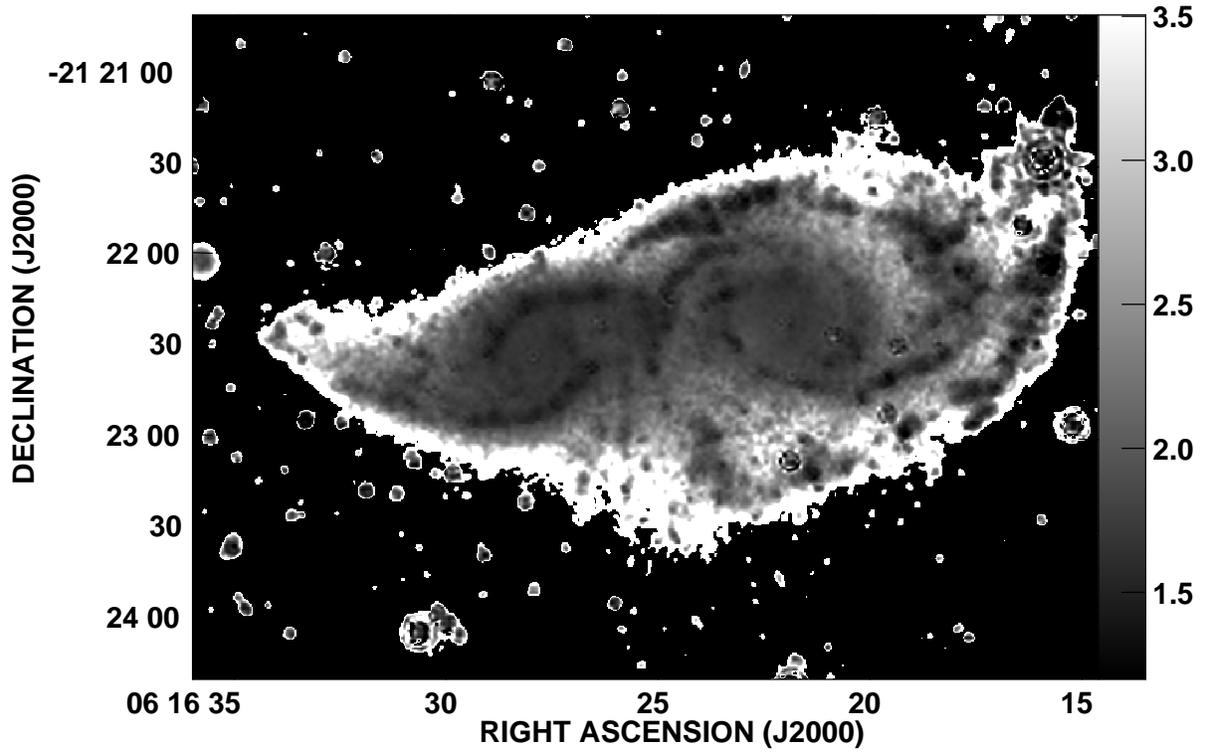} \caption{A map of 3.6 $\mu$m
divided by the 4.5 $\mu$m. Black
regions have low values of this ratio. The nuclei disappear but
the star-forming beads become more prominent as dark regions.}
\label{fig:divide}\end{figure}

\begin{figure}
\epsscale{1.0} \plotone{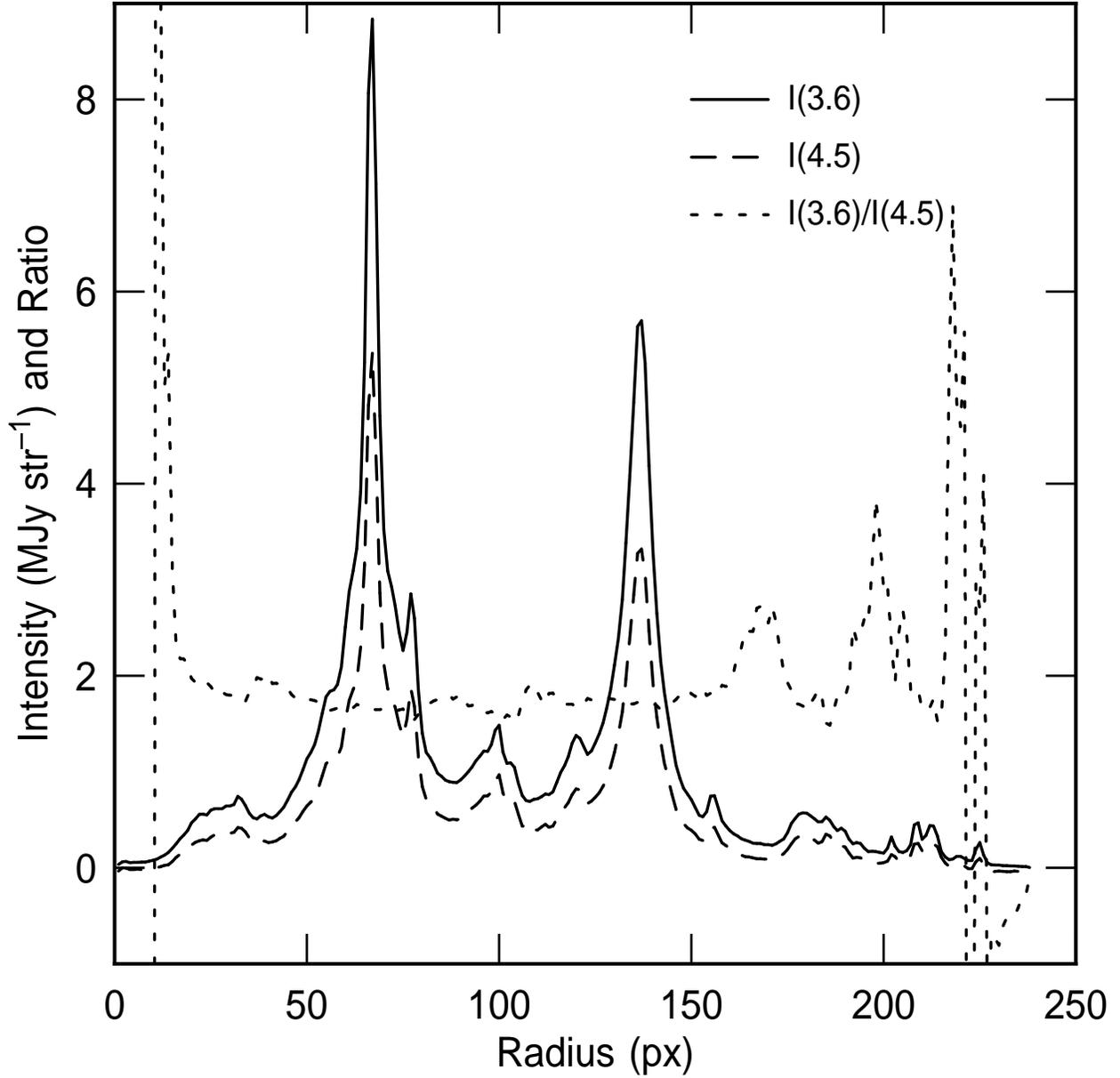} \caption{Radial profiles of
intensity for $3.6\mu$m  and $4.5\mu$m emission and of the
I(3.6)/I(4.5) ratio show the nuclei of both galaxies as the
biggest peaks in the individual channels. The radial color profile
is uniform in the interarm and nuclear regions and dips down in
spiral arms.} \label{fig:cprofile}\end{figure}

\begin{figure}
\epsscale{1.0} \plotone{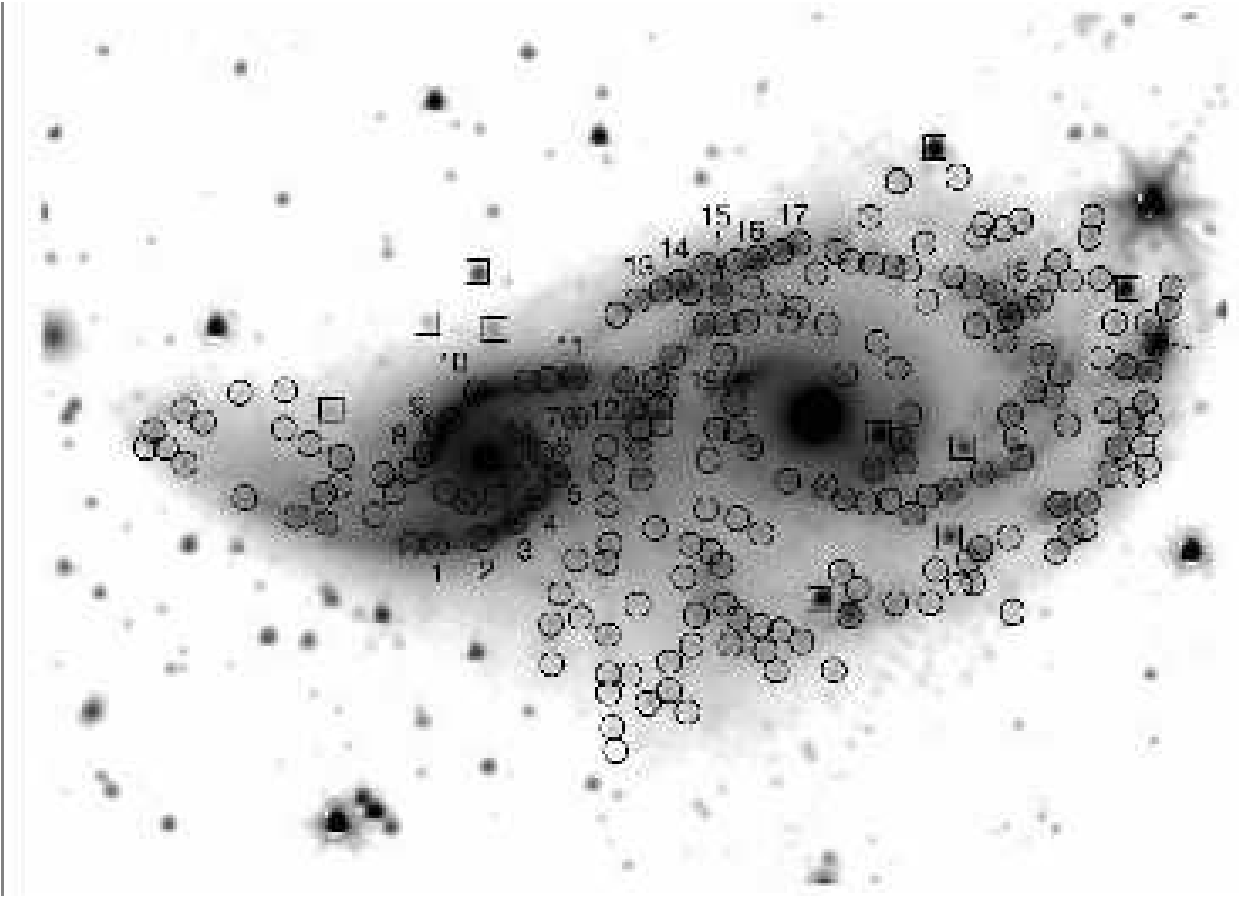} \caption{Measured clumps are shown
circled on this IRAC $3.6\mu$m image; the objects in squares are
stars. The image has a logarithmic intensity scale. The clumps
whose properties are in Table 2 are indicated by numbers. Feature
i is also labeled. A total of 225 clumps were measured. (Image degraded for astroph)}
\label{fig:id}\end{figure}

\begin{figure}
\epsscale{0.7} \plotone{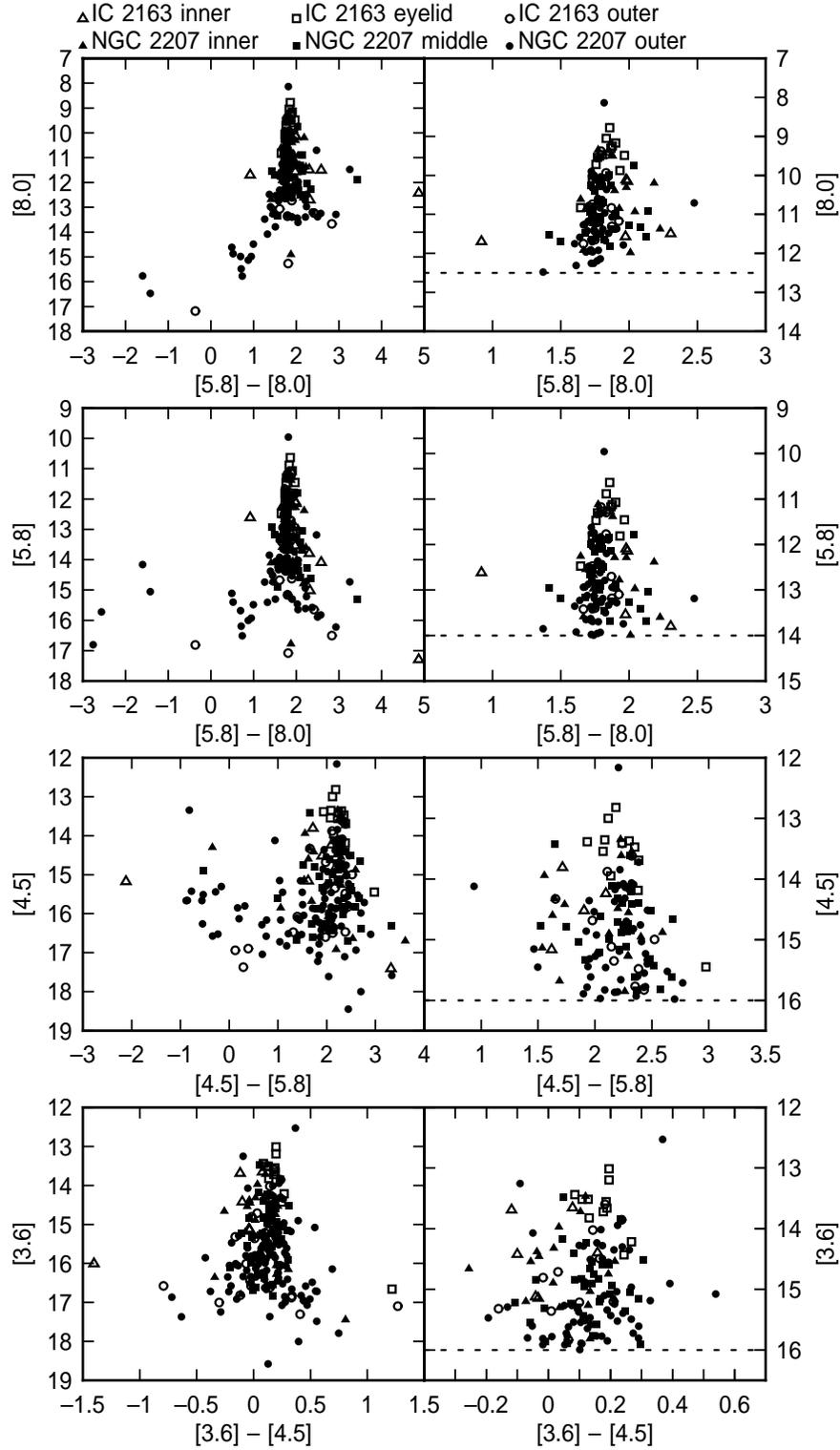} \caption{Color-magnitude diagrams
are shown subdivided according to different positions in the two
galaxies, with open symbols for clumps in IC 2163 and closed
symbols for clumps in NGC 2207. The left-hand panels include all
clumps, whereas the right-hand panels show only the low-noise
clumps. Horizontal dotted lines are the estimated sensitivity
limits. Feature i is the brightest source, followed by clumps in
the eyelid of IC 2163.} \label{fig:cm}\end{figure}

\begin{figure}
\epsscale{1.0} \plotone{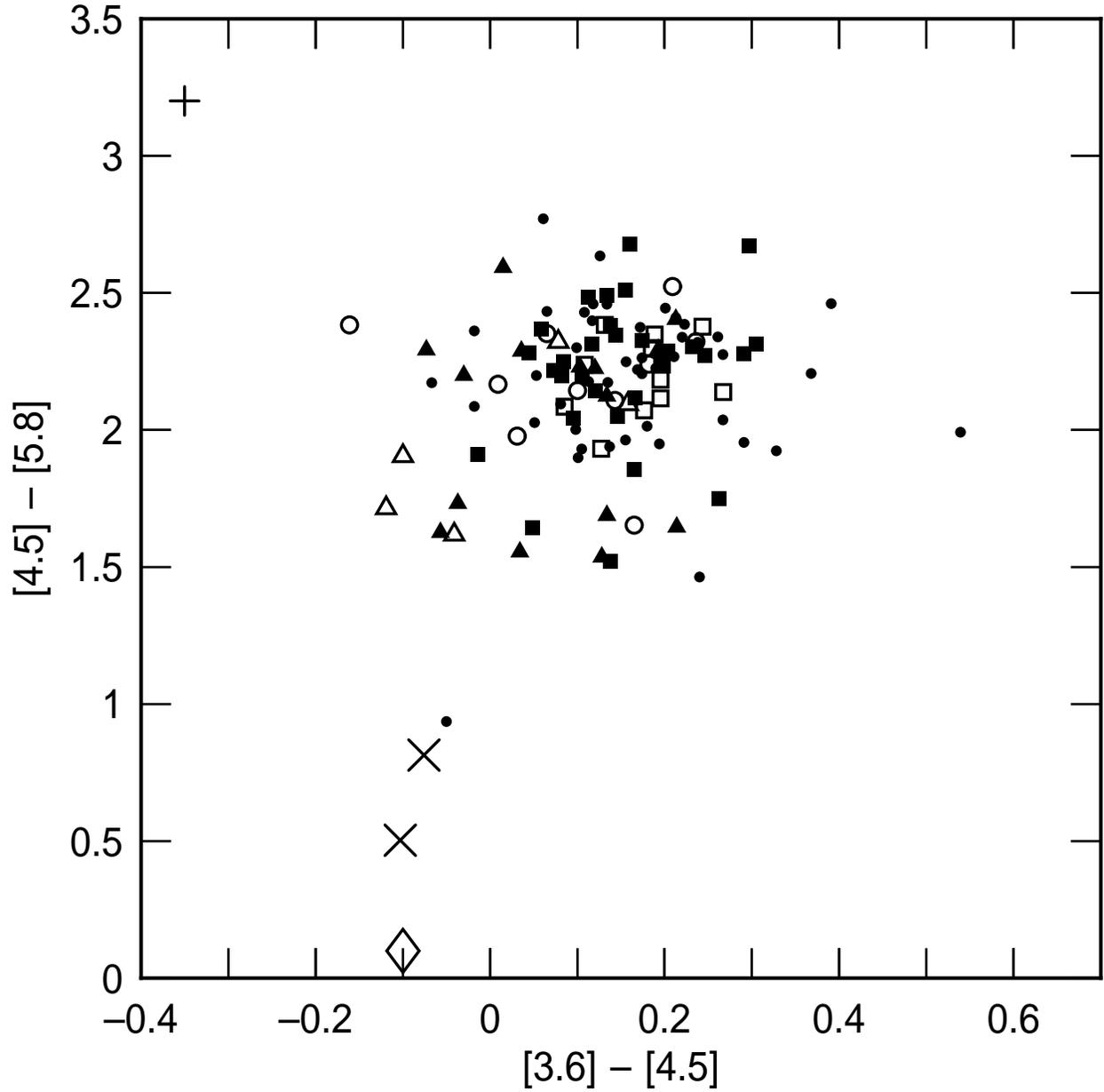} \caption{Color-color diagrams for
[4.5]-[5.8] versus [3.6]-[4.5] separated according to regions in
IC 2163 and NGC 2207, with the same symbols as in the previous
figure. The diamond represents late-type stars, the plus sign
represents dust, and the two X's represent the nuclei of IC 2163
and NGC 2207, which are predominantly starlight in the IR. }
\label{fig:cc123}\end{figure}

\begin{figure}
\epsscale{1.0} \plotone{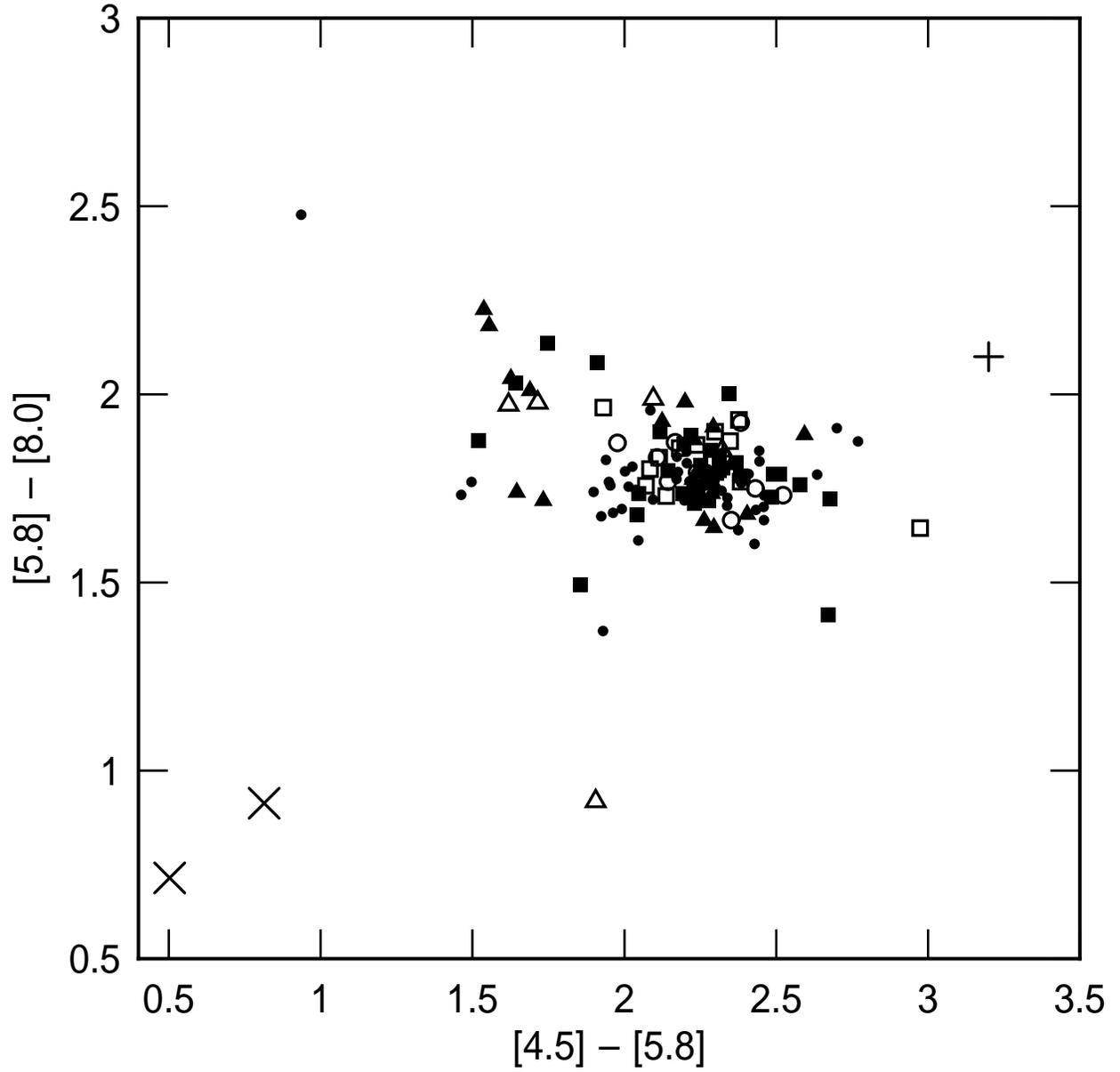} \caption{Color-color diagrams for
[5.8]-[8] versus [4.5]-[5.8], with the same symbols as in the
previous figure. } \label{fig:cc234}\end{figure}

\begin{figure}\epsscale{1.0}\plottwo{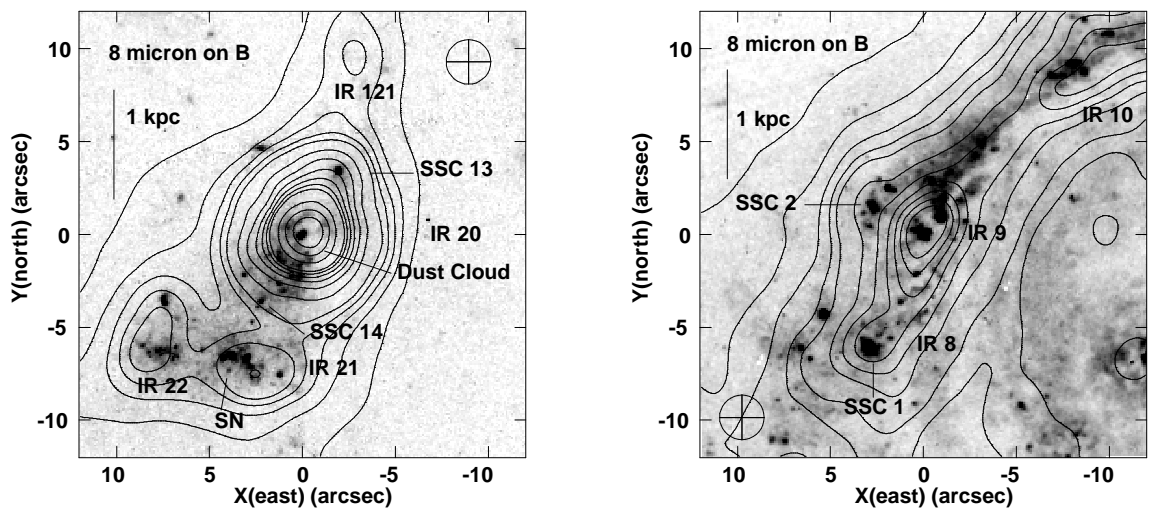}{f10b.eps}
\caption{Two regions of star formation shown with spiral IRAC
$8\mu$m contours on HST B-band grayscale. The contours are
2...(2)...10...(5)...40, 60, 80, 100 MJy/ster and the field of
views are 24 arcsec on a side, like the integrated regions in
Table 1. Feature i (left) is the brightest source in both
galaxies. It contains two super star clusters, numerous other
clusters and associations, 4 IRAC sources, supernova
1999ec, and several dark dust clouds. IRAC source 9 (right) is in
the eyelid of IC 2163. The field of view contains 2 super star
clusters and 2 other IRAC sources. The 2.2 arcsec resolution at $8\mu$m is
shown by the circle with a plus-sign.}
\label{fig:examples}\end{figure}

\begin{figure}
\epsscale{1.0} \plotone{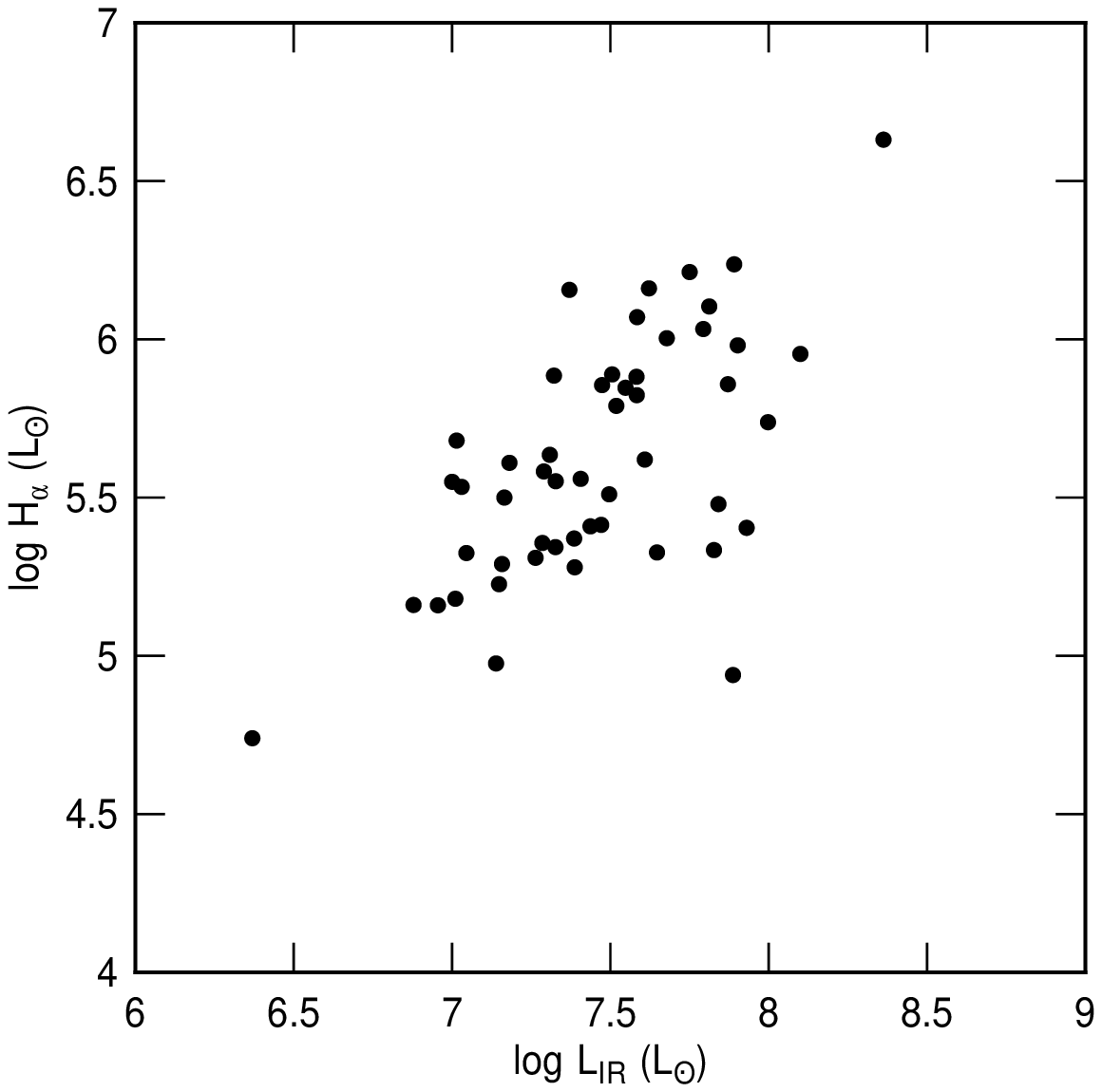} \caption{The H$\alpha$ and IRAC
luminosities for the clumpy emission are shown. The correlation is
approximately linear. The wide scatter is presumably the result of
variable ages and extinction in the optical.}
\label{fig:halpha}\end{figure}

\begin{figure}
\epsscale{1.0} \plotone{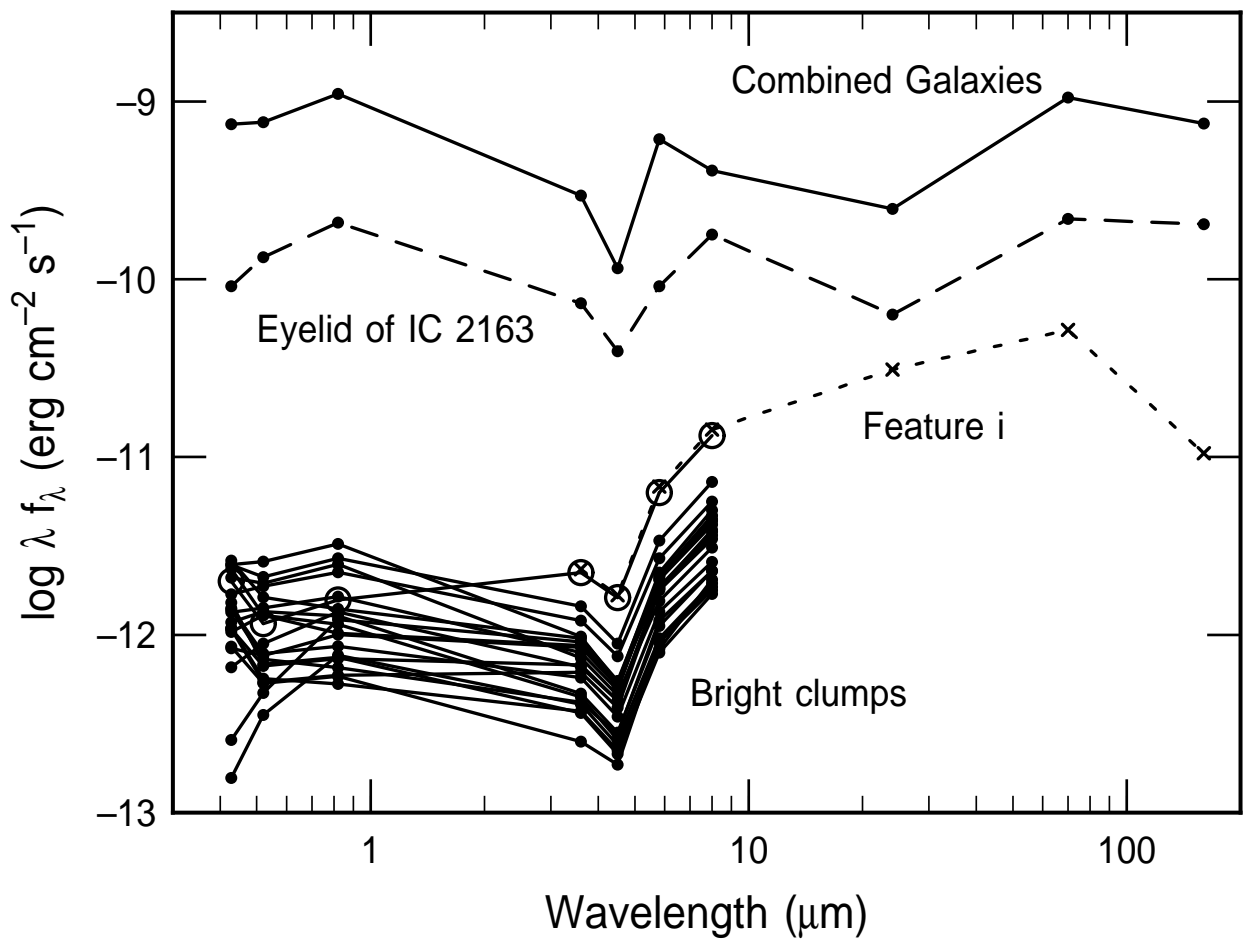} \caption{(solid lines) Spectral
energy distributions are shown for the clumps in Table 2, with the
log of the flux density, $\lambda F_\lambda$, plotted as a
function of the wavelength. HST and SST data are combined. The
SEDs are also shown for the combined galaxies, the eyelid of IC
2163, Feature i at short wavelength (line with open circles) and a
$24^{\prime\prime}$ box around Feature i at all wavelengths
(dotted line with x-marks). } \label{fig:sedhubble}\end{figure}

\begin{figure}
\epsscale{.8} \plotone{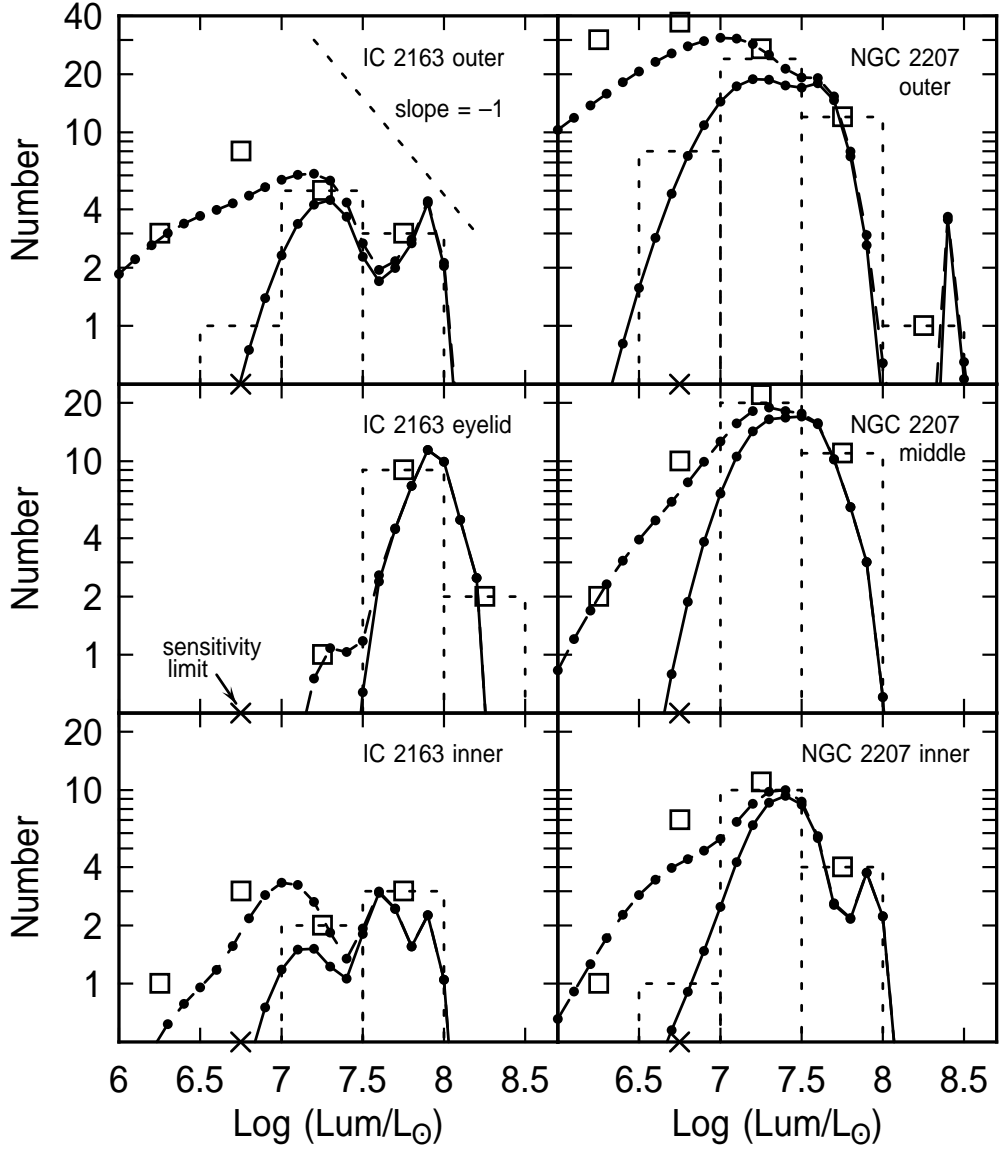} \caption{The luminosity
distribution function based on IRAC emission is shown for the
clumps in IC 2163 and NGC 2207, sorted according to location. The
brightest region in both galaxies (Feature i) is the single peak
to the right of the main peak in the upper right panel. The dotted
histograms indicate counts of low-noise clumps in bins of 0.5 in
the log of the luminosity. The open squares are counts for all of
the clumps, binned in the same way. The curves are histograms of
another type, binned in intervals of 0.1 in the log, and made by
summing Gaussian contributions from each clump. This smooths out
the luminosity functions and allows inclusion of the measurement
uncertainties as dispersions in the Gaussians. The solid line is
for the low-noise clumps and the dashed line is for all of the
clumps, including those whose luminosities are not well
determined. The x-marks on the abscissae are the estimated
sensitivity limits, based on the upper limits to the magnitudes on
the right-hand side of Fig. 7. The dotted line in the upper left
panel has a slope of $-1$, which is typical for cluster mass
functions.} \label{fig:gaush}\end{figure}

\begin{figure}
\epsscale{1.0} \plotone{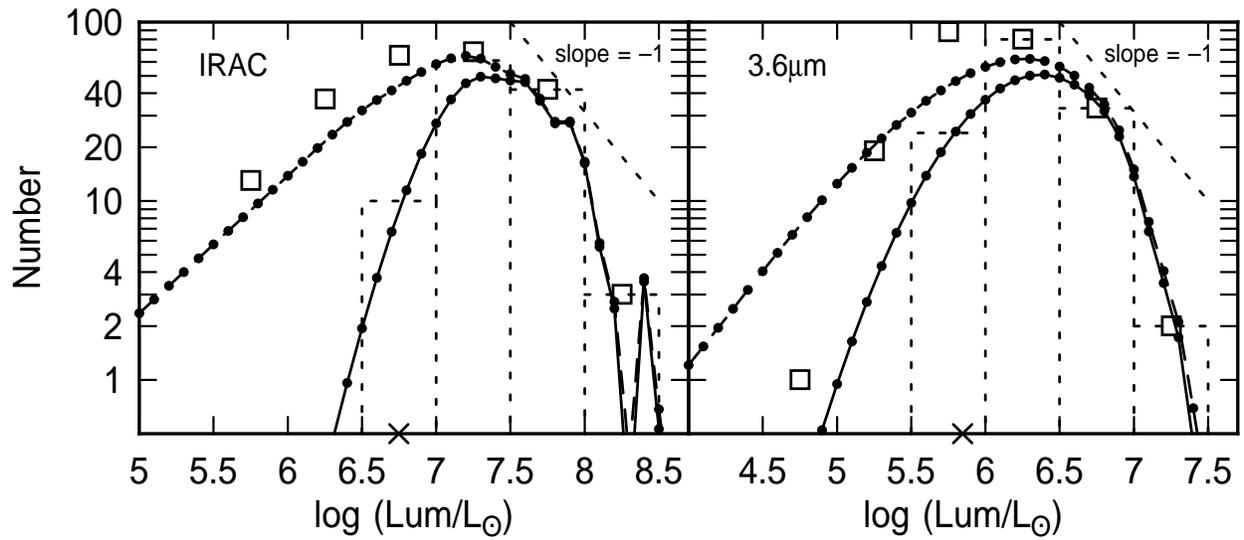} \caption{The luminosity
distribution functions for IRAC (left panel) and 3.6$\mu$m (right
panel) bands, regardless of location. The open squares are counts
for all of the clumps in bins of 0.5 in the log of luminosity. The
curves are Gaussian sums as in Fig. 13 with solid lines for low-noise
clumps and dashed lines for all the clumps. The bins used for the
Gaussian sums are shown by the dots.
Sensitivity limits for all IRAC bands combined and for the
$3.6\mu$m band alone are plotted as x-marks on the abscissae.}
\label{fig:gausa}\end{figure}

\begin{figure}
\epsscale{1.0} \plotone{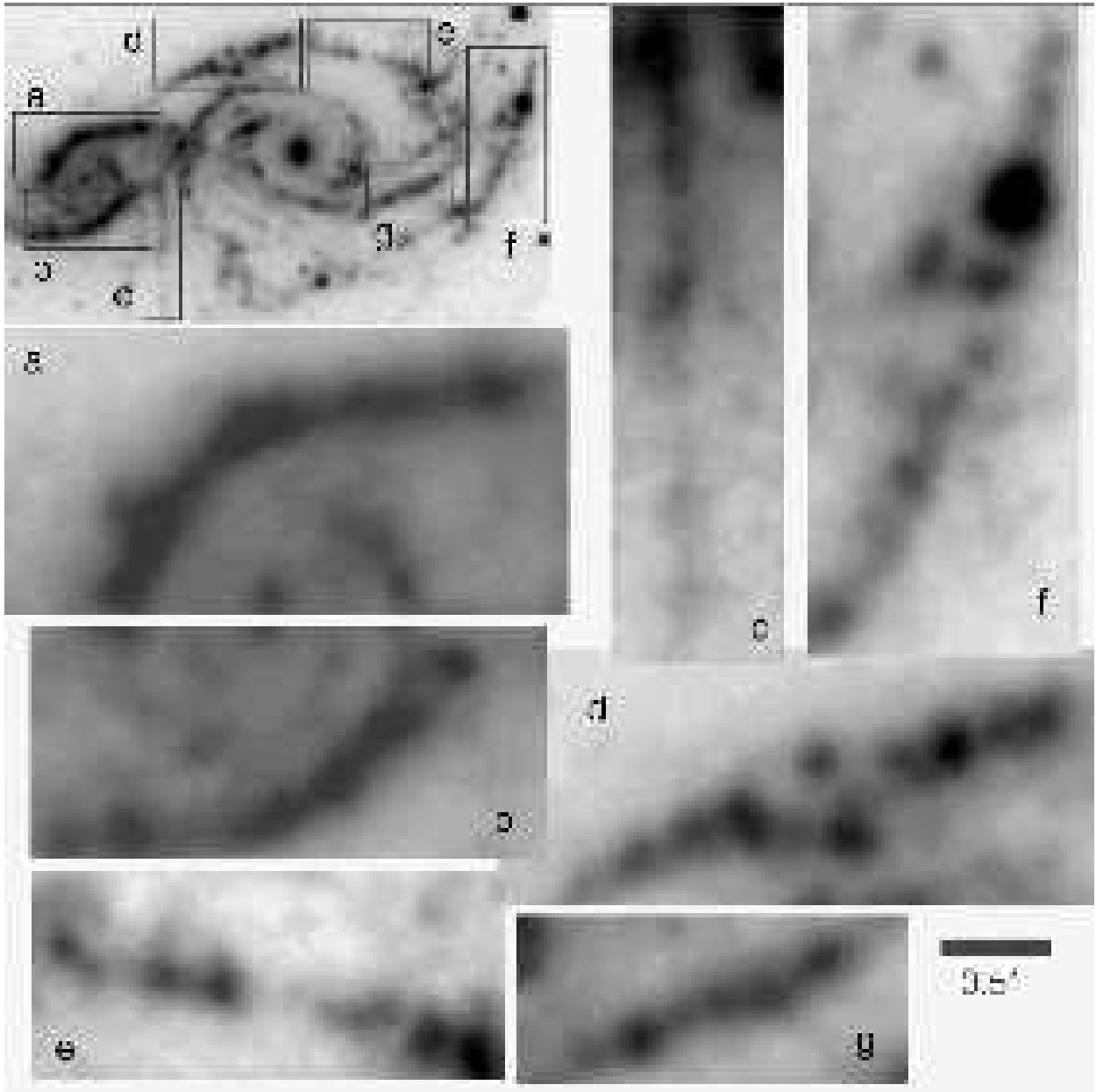} \caption{The 8$\mu$m logarithmic
image is shown in the upper left, with outlined strings of clumps
shown enlarged below. The scale of 0.5 arcmin applies to the
enlarged figures and corresponds to $\sim
5$ kpc. Virtually all of the bright clumps are in such strings.
The regularity of the strings may account for the non-power law
luminosity functions. (Image degraded for astroph)} \label{fig:beads}\end{figure}

\begin{figure}
\epsscale{1.0} \plotone{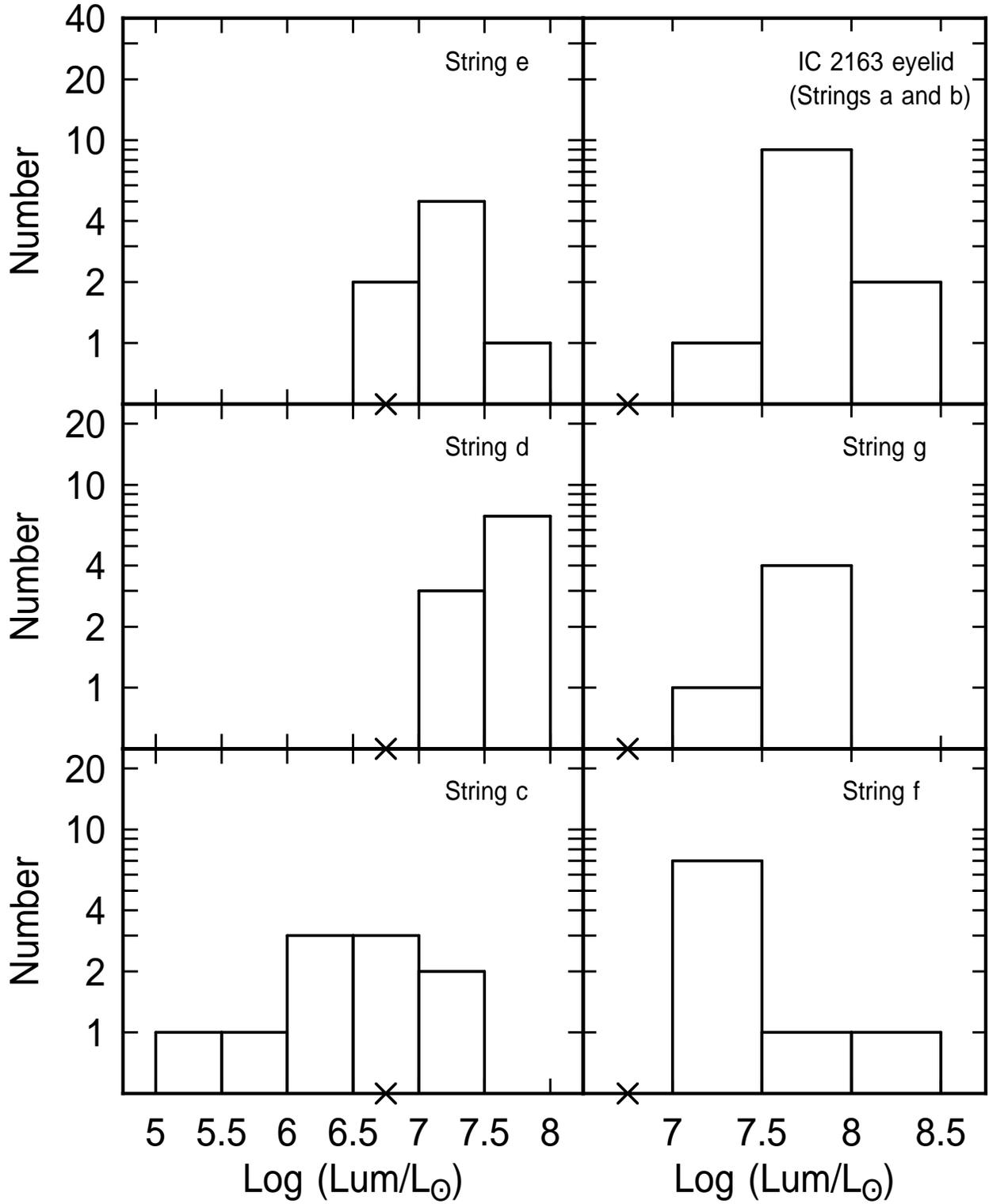} \caption{Histograms for the
distributions of IRAC luminosities in each of the 6
bead-on-a-string regions shown in Fig. 15. The detection limit is
indicated by an x-mark on the abscissa of each panel. }
\label{fig:beadhis}\end{figure}

\begin{figure}
\epsscale{.8} \plotone{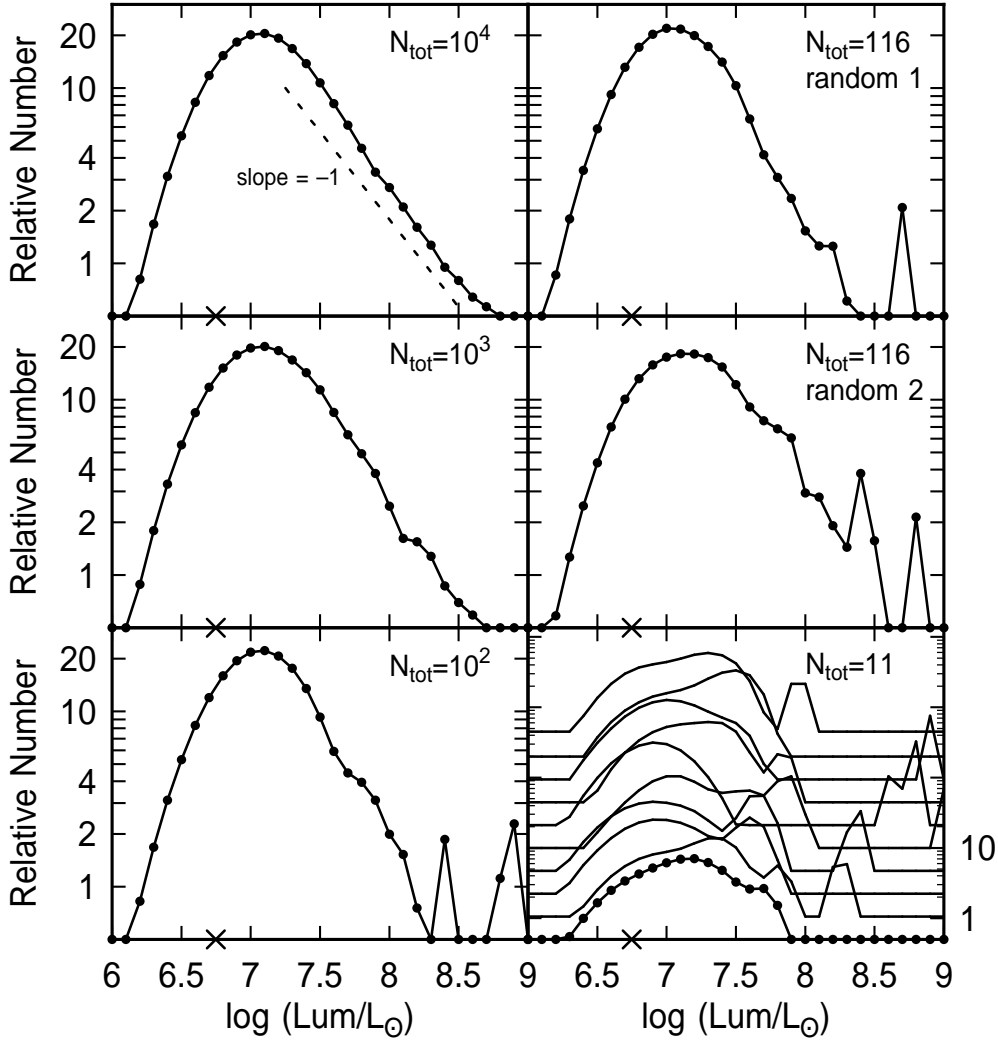} \caption{Model luminosity
functions of a coeval population of clumps with an intrinsic power
law mass function having a slope of -1 on this diagram and a lower
luminosity cut-off of log L=6.75, the sensitivity limit found for
IC2163/NGC 2207 (shown by the x-mark on each abscissa). The models
assume random sampling from the intrinsic function. The
Gaussian-sum technique is used to smooth the curves and to take
into account the effect of measurement errors dependent on
luminosity, as in Figs. 13-14. The plotted numbers are normalized
to give the same histogram areas in each panel.  The actual
numbers of clumps used for these models are indicated. The top two
right-hand panels use different random numbers but both have a
number of clumps equal to the observed number of low-noise clumps
in our galaxy pair, 116. The bottom right panel has 10 random
trials with only 11 clumps each, the same number as in the eyelid.
Each trial is shifted upward for clarity; the vertical scale is
indicated on the right hand axis. These stochastic models show
that the overall power law should be apparent even for as few as
$\sim 100$ clumps. Fluctuations at the high luminosity end for the
$N\sim100-116$ cases can account for stray peaks such as that
produced in Fig. 13 by Feature i. When $N$ is very low, such as
11, the model functions still have a peak near the sensitivity
limit, unlike the eyelid region. } \label{fig:gausm}\end{figure}

\begin{figure}
\epsscale{1.0} \plotone{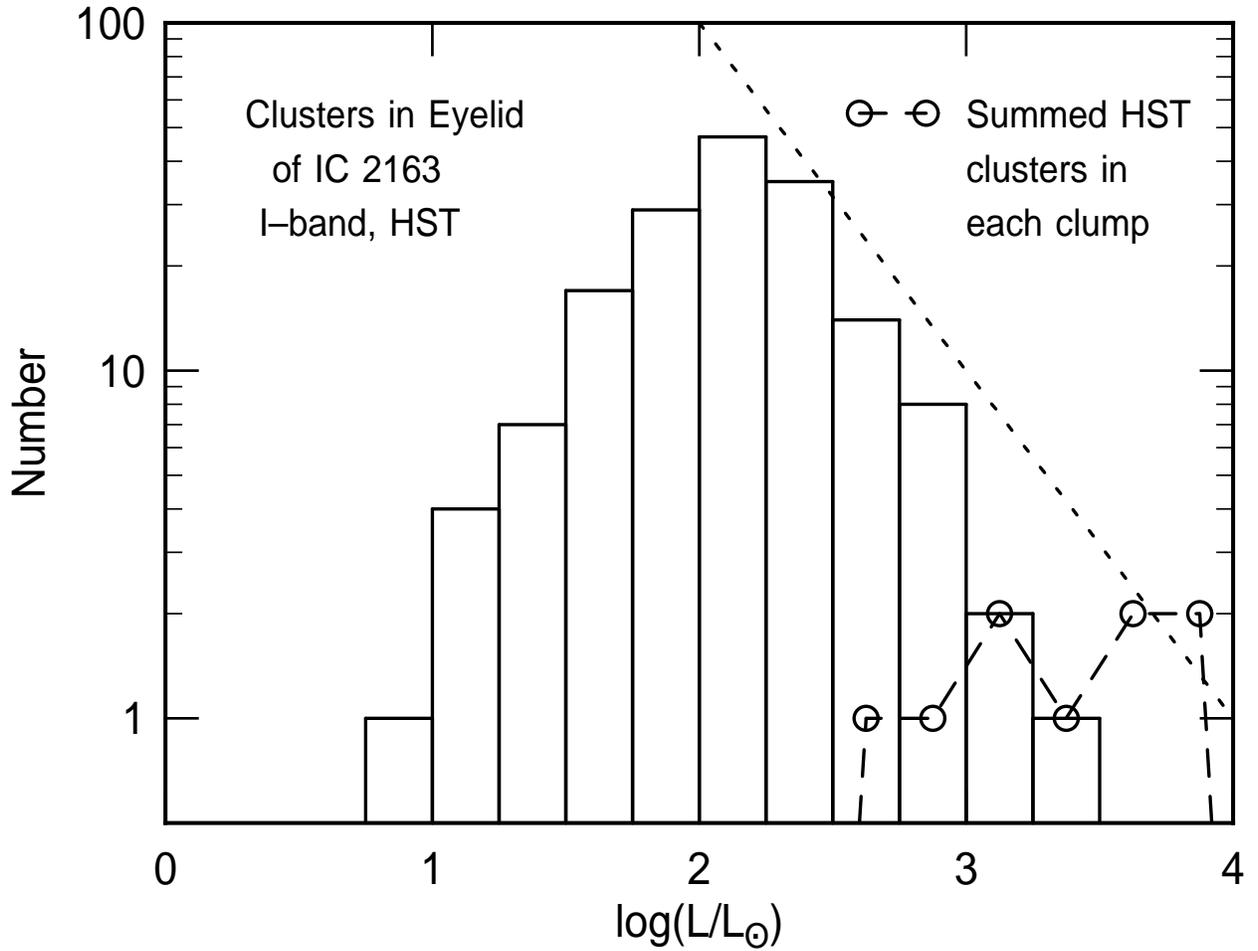} \caption{Luminosity distribution
function in the I-band of 165 star clusters observed by HST in the
eyelid region of IC 2163. The dashed line with circles is the
luminosity function at I-band for the sum of the clusters in each
IRAC-defined star complex. The dotted line has a slope of $-1$. }
\label{fig:hst}\end{figure}

\end{document}